\documentclass{aa}  
\usepackage{natbib}
\usepackage{graphicx}
\usepackage{txfonts}
\begin{document}

   \title{The undetectable fraction of core-collapse supernovae in luminous infrared galaxies}
    \subtitle{II. GSAOI/GeMS dataset}

   \author{I. Mäntynen\inst{1}\fnmsep\thanks{Corresponding author; \email{iamant@utu.fi}}
          \and
          E. Kankare
          \inst{1}
          \and
          S. Mattila
          \inst{1, 2}
          \and
          A. Efstathiou
          \inst{2}
          \and
          S. D. Ryder
          \inst{3,4}
          \and
          E. Kool
          \and
          K. Matilainen
          \inst{1}
          \and
          T. M. Reynolds
          \inst{1,5,6}
          \and
          C. Vassallo
          \inst{1}
          \and
          P. Väisänen
          \inst{7,8}
          }

   \institute{Tuorla Observatory, Department of Physics and Astronomy, University of Turku, 20014 Turku, Finland
    \and
    School of Sciences, European University Cyprus, Diogenes street, Engomi, 1516 Nicosia, Cyprus
    \and
    School of Mathematical and Physical Sciences, Macquarie University, Sydney, NSW 2109, Australia
    \and
    Astrophysics and Space Technologies Research Centre, Macquarie University, Sydney, NSW 2109, Australia
    \and
    Cosmic Dawn Center (DAWN)
    \and
    Niels Bohr Institute, University of Copenhagen, Jagtvej 128, 2200 København N, Denmark
    \and
    Finnish Centre for Astronomy with ESO (FINCA), University of Turku, 20014 Turku, Finland
    \and
    South African Astronomical Observatory, P.O. Box 9, Observatory, 7935 Cape Town, South Africa
             }

   \date{Received September 15, 1996; accepted March 16, 1997}

  \abstract 
   {Core-collapse supernovae (CCSNe) in luminous infrared galaxies (LIRGs) can have extreme line-of-sight host galaxy dust extinctions, which leads to a large fraction of the events remaining undetected by optical and infrared surveys. This population of undetected CCSNe is important to constrain in order to determine the cosmic CCSN rates, which can be used to estimate the cosmic star formation history independently from methods based on galaxy luminosities.}
   {Our aim is to confirm and refine our estimates for the undetectable fraction of CCSNe in LIRGs in the local Universe. Our study is based on the near-infrared \textit{K}-band multi-epoch SUNBIRD survey monitoring dataset of a sample of nine LIRGs using the Gemini-South telescope with the multi-conjugate GSAOI/GeMS laser guide star adaptive optics system.}
   {We determined the limiting magnitudes for CCSN detection for each epoch in our dataset with artificial supernova injection and image subtraction methods. Subsequently, we used a Monte Carlo method to determine the combined effects of limiting magnitudes, survey cadence, CCSN subtype distribution, and their light curve evolution diversity. The intrinsic CCSN rates of the sample galaxies were estimated based on detailed modelling of their spectral energy distribution. Finally, we combined the resulting CCSN detection probabilities with the intrinsic CCSN rates for the dataset, and compared that against the real CCSN detections over the survey period. }
   {Based on our GSAOI/GeMS dataset, assuming optical or near-infrared example surveys with capabilities to detect CCSNe in local LIRGs with host extinctions of $A_V =$ 3 or 16 mag, respectively, the resulting total undetectable fractions are $86.0^{+4.7}_{-5.9}$ \% and $53.6^{+15.6}_{-19.6}$ \%. When folding in the results from our previous near-infrared adaptive optics assisted LIRG monitoring dataset, the corresponding total undetectable fractions are $88.3^{+2.6}_{-3.2}$ \% and $61.4^{+8.5}_{-10.6}$ \%, respectively.}
   {}


   \maketitle

\section{Introduction}
\label{introduction}

    Core-collapse supernovae (\mbox{CCSNe}) are the terminal explosions at the end of the lifetime of massive stars above $\sim$8 M$_{\odot}$ (e.g., \citealt{CCSN_prog_mass}). Due to their relatively short lifespans, \mbox{CCSNe} can be used to trace the star formation rate (\mbox{SFR}). Estimating the SFR from the CCSN rate ($\mathrm{SFR_C}$) is a method  independent from the usual luminosity-based methods ($\mathrm{SFR_L}$). There has been an ongoing discussion whether the $\mathrm{SFR_C}$ and the $\mathrm{SFR_L}$ agree at different redshift ranges. \cite{horiuchi2011} noted a discrepancy between the reported $\mathrm{SFR_C}$ and $\mathrm{SFR_L}$ values, with the latter being roughly two times larger than the former at $0 < z < 0.4$. This was challenged by \cite{cappellaro2015}, who found the $\mathrm{SFR_C}$ and $\mathrm{SFR_L}$ to be in agreements within errors for $0.05 < z < 0.35$. \cite{botticella2012} noted that the choice of the observational wavelength region in determining the $\mathrm{SFR_L}$ has a large impact; for example, SFRs determined from H$\alpha$ were found to be lower by a factor of about two compared to SFRs determined from far ultraviolet (FUV). Similarly \cite{strolger2015} found a difference in $\mathrm{SFR_L}$ obtained from UV plus IR observations and IR-only observations, with the UV plus IR derived $\mathrm{SFR_L}$ matching the $\mathrm{SFR_C}$. \cite{gruppioni2020} comprehensively collated different SFR estimates obtained through various approaches. The estimates are in good agreement at low \textit{z}, but rapidly diverge after $z\approx1$. This is expected to some extent due to the inherent difficulties in observing at larger redshifts and is further enhanced by the increased contribution of dust-obscured star formation at higher redshifts \citep{magnelli2011,zavala2021}. In particular, CCSNe that remain undetected by surveys due to dust-obscuration have a major effect on the $\mathrm{SFR_C}$ estimate \citep{mattila2012}.

    Luminous infrared galaxies (LIRGs) are defined as galaxies with $ 10^{11}L_{\odot} \leq L_{\mathrm{IR}} < 10^{12}L_{\odot}$. The dominant source of the infrared luminosity is the reprocessing of UV and optical photons by dust and gas \citep{Kennicutt1998,torres2021}. The source of the high-energy photons is hot young stars, active galactic nuclei (AGNs), or a combination of both. Luminous infrared galaxies are often in the process of merging or show signs of past interaction. Star formation in LIRGs is concentrated in the nuclear and circumnuclear regions at $\sim 0.1$ to $\sim 1$ kpc scales \citep{soifer2001}, which results in a substantial fraction of CCSNe remaining undetected due to extinction and insufficient spatial resolution \citep{kool2018}. While relatively rare in the local Universe, the contribution of LIRGs to the cosmic SFR becomes more significant at higher redshifts. At \textit{z} = 1 the contribution of LIRGs to the total SFR at that redshift exceeds 50\% \citep{magnelli2009}. Thus, the study of cosmic SFR beyond the local Universe with CCSNe requires detailed understanding of these events in LIRGs.
    
    \cite{mattila2012} found that the fraction of undetected CCSNe in LIRGs by optical surveys can be as high as $83^{+9}_{-15}\%$. This result was based on 14 years of reported observations of the LIRG Arp 299 by various transient surveys, where the estimated intrinsic CCSN rate was compared to the number of detected events.
    \cite{miluzio2013} found that by shifting a supernova (SN) survey to the \textit{K}-band (where the extinction in flux is only $\sim10^{-4}$ of optical) $\sim$60 to 75\% of the CCSNe in LIRGs are not detected. Their result was based on detection efficiencies of artificial SNe simulated on the sample galaxies. 
    \cite{fox2021} used the Spitzer Space Telescope at 3.6 $\mu$m to survey high SFR galaxies for CCSNe. Despite the advantages of a space-based telescope operating in the near-IR, their results suggest that $\sim$83\% of CCSNe remained undetected at <150 Mpc. This likely resulted from the complex point spread function (PSF) of Spitzer data, resulting in poor detection efficiency in the central regions of galaxies, where most of the CCSNe explode. 
    \citeauthor{paper1} (2025; hereafter Paper I) find that  $66.0^{+8.6}_{-14.6}$\% and $89.7^{+2.6}_{-4.4}$\% of CCSNe in LIRGs in the local Universe are not detected by near-IR (assuming an extinction threshold of $A_V$ = 16 mag) and optical ($A_V$ = 3 mag) surveys, respectively. The result is based on adaptive optics (AO) monitoring of a sample of eight LIRGs using the Gemini-North telescope.
    
    While the reduced impact of host extinction with infrared observations is quite effective compared to optical observations, a higher resolution is required to further lower the percentage of missed CCSNe. \cite{kool2018} show that supplementing the advantages of the \textit{K}-band with high angular resolution AO observations can decrease the fraction of undetected CCSNe in LIRGs, especially in the circumnuclear regions. Several CCSNe in central regions of LIRGs have been detected using this technique (e.g. \citealt{mattila2007,kankare2008,kankare2012}). 
    Radio observations are unaffected by dust obscuration, and radio interferometry can reach extremely high angular resolutions. Unfortunately, only a fraction of CCSNe are bright enough in radio wavelengths to be readily detectable (e.g. \citealt{romero2014}). \cite{varenius2019} estimate that only $\sim$2.5 to $\sim$5.6 \% of CCSNe in the nearby ultraluminous infrared galaxy (ULIRG; $10^{12} L_{\odot} \leq L_{\mathrm{IR}} \leq  10^{13} L_{\odot}$) Arp 220 at 87 Mpc were detectable in radio.
    Thus, the high angular resolution surveys at near-IR bands provide the most comprehensive view of CCSNe in LIRGs, which in turn allows for a more accurate matching of $\mathrm{SFR_C}$ and $\mathrm{SFR_L}$ in the redshift regime, where most of the star formation is obscured by dust.

    Studies of CCSN rates typically apply an extinction correction to their observed rates to take into account the effect of host extinction (e.g. \citealt{dahlen2012}). However, extinction correction models for normal galaxies primarily suggest moderate obscuration and are not sufficient to explain the observed CCSN population in LIRGs, where the majority of these events can remain undetected. Furthermore, in addition to a population of CCSNe in LIRGs with host extinctions that could be described with a more obscured extinction correction model, there is a population of CCSNe in these galaxies with extremely high host extinctions, which remain undetected in all optical or infrared observations; this population is called the missing fraction \citep{mattila2012}. The CCSNe within the missing fraction explode deep in their host galaxies in or behind dusty molecular clouds, primarily in the central regions of LIRGs. \cite{MM01} measured extinctions of 19 SN remnants from the nearby starburst galaxy M~82 in the central 500 pc region based on H I and $\mathrm{H_2}$ column density measurements. This resulted in a mean extinction of $A_{V} = 24 \pm9$ mag. Contrasting this with an example model of host extinction distribution for normal spiral galaxies by \cite{riello2005}, excluding edge-on cases, the mean extinction barely reaches $A_{V} = 0.5$ mag. Unfortunately, there is no host extinction model for CCSNe in LIRGs that would be based on a large number of SNe. However, in Paper I an extinction correction was derived based on a sample of 19 CCSNe in LIRGs \citep{kankare2021}, which we adopted in our analysis.

    Classical CCSN classification divides the SNe into hydrogen-rich Type II and hydrogen-poor Type I (also known as stripped-envelope SNe) based on their spectra. H-poor SNe are divided into Types IIb, Ib, and Ic. Spectroscopically they show progressively increased levels of envelope stripping, respectively showing some hydrogen, helium but no hydrogen, and neither hydrogen nor helium. Additionally, many subtypes can be further spectroscopically classified into narrow, broad-line, and peculiar subtypes. For example, Type IIn SNe are H-rich events that show narrow spectral lines ($\lesssim 1000$ km s$^{-1}$). SN 1987A-like events are peculiar Type II SNe that resemble the historical SN 1987A, with a distinctive slowly rising light curve.

    Here we build upon the results from Paper I by including observations conducted with the Gemini South telescope. Our methodology is the same as in Paper I: We performed a standard data reduction (Sect. \ref{data_set}), carried out algorithmically optimised template image subtraction (Sect. \ref{automatic_image_subtraction}), determined the limiting magnitudes of the dataset via artificial SN injection (Sect. \ref{limiting_magnitudes}), and used a Monte Carlo simulation to determine the effects of different CCSN subtypes, peak magnitude variation, evolution timescale variability, host extinction, and survey cadence to produce CCSN detection probabilities for the dataset (Sect. \ref{monte_carlo_method}). Finally, the intrinsic CCSN rates for the dataset were obtained via spectral energy distribution (SED) analysis through Markov Chains (SATMC). These were compared against the real detected CCSNe in the dataset to determine the missing fraction. Additionally, we estimated the fraction of CCSNe that remain undetected due to more moderate extinctions, in addition to the missing fraction. We calculated this fraction from a host extinction model. These two fractions form the total undetectable fraction. Results are summarised in Sect. \ref{results}, and we briefly discuss the implications of our results on the CCSN detectability at different redshifts in Sect. \ref{discussion}. Conclusions are provided in Sect. \ref{conclusions}.

\section{Dataset}
\label{data_set}

    The dataset used in this study was obtained by our Supernovae UNmasked By Infra-Red Detection (SUNBIRD) collaboration \citep{kool2018}. The collaboration has used multiple AO programmes to discover and study CCSNe in LIRGs (e.g. \citealt{kankare2012,kankare2014}) and characterise the super-star cluster population in these galaxies (e.g. \citealt{zara2013,zara2019,zara2022}). 
    
\begin{figure*}
\centering
\includegraphics[width=0.33\linewidth]{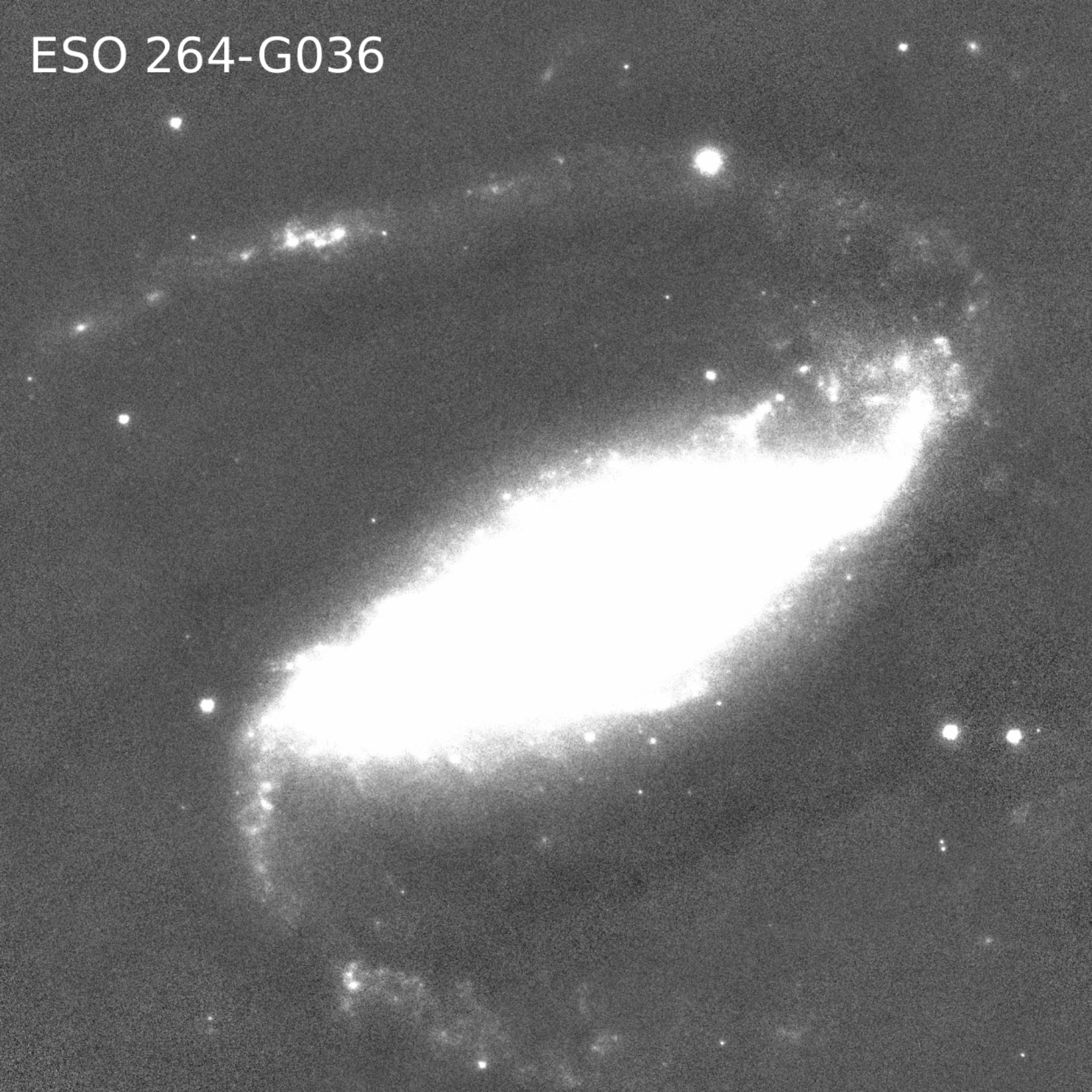}
\includegraphics[width=0.33\linewidth]{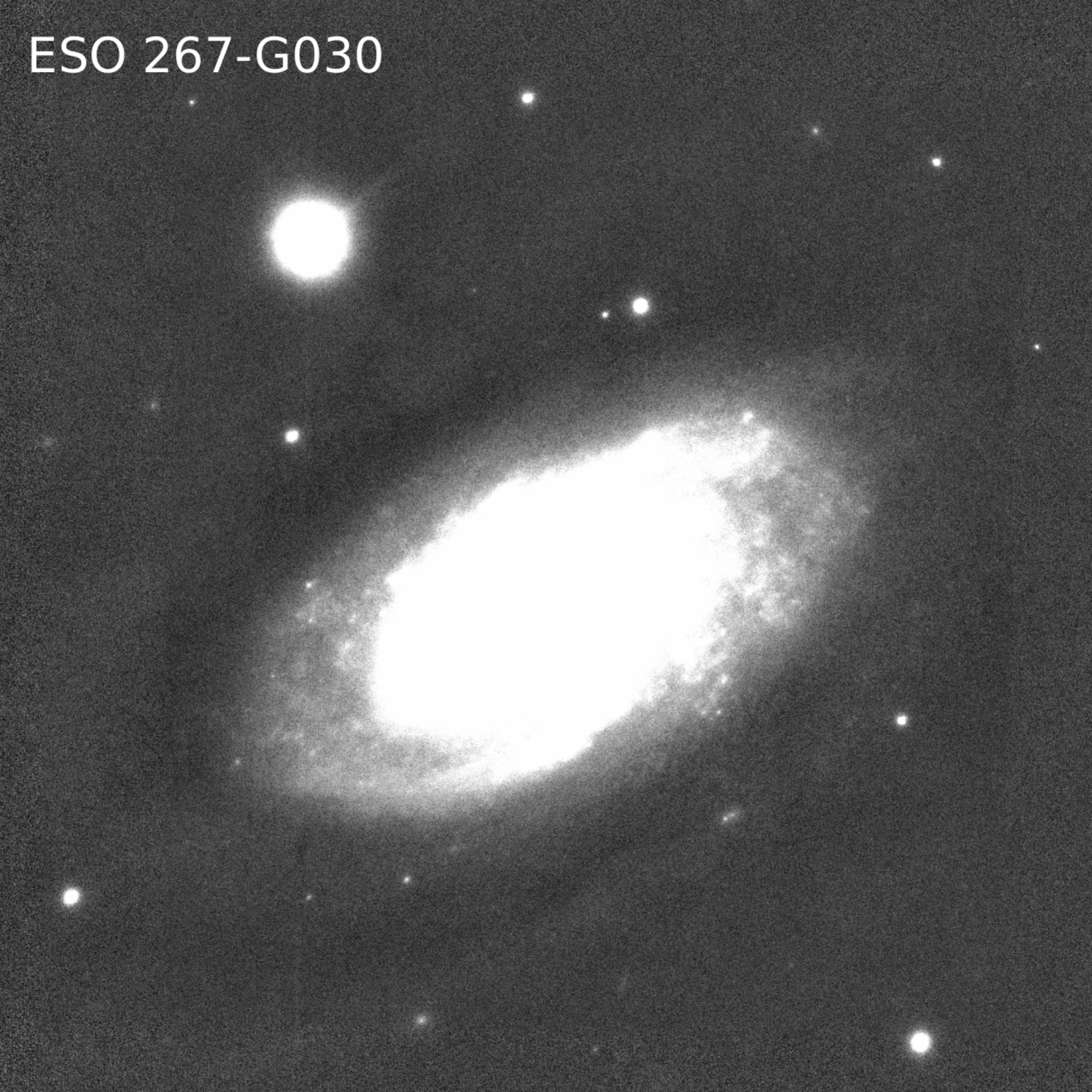}
\includegraphics[width=0.33\linewidth]{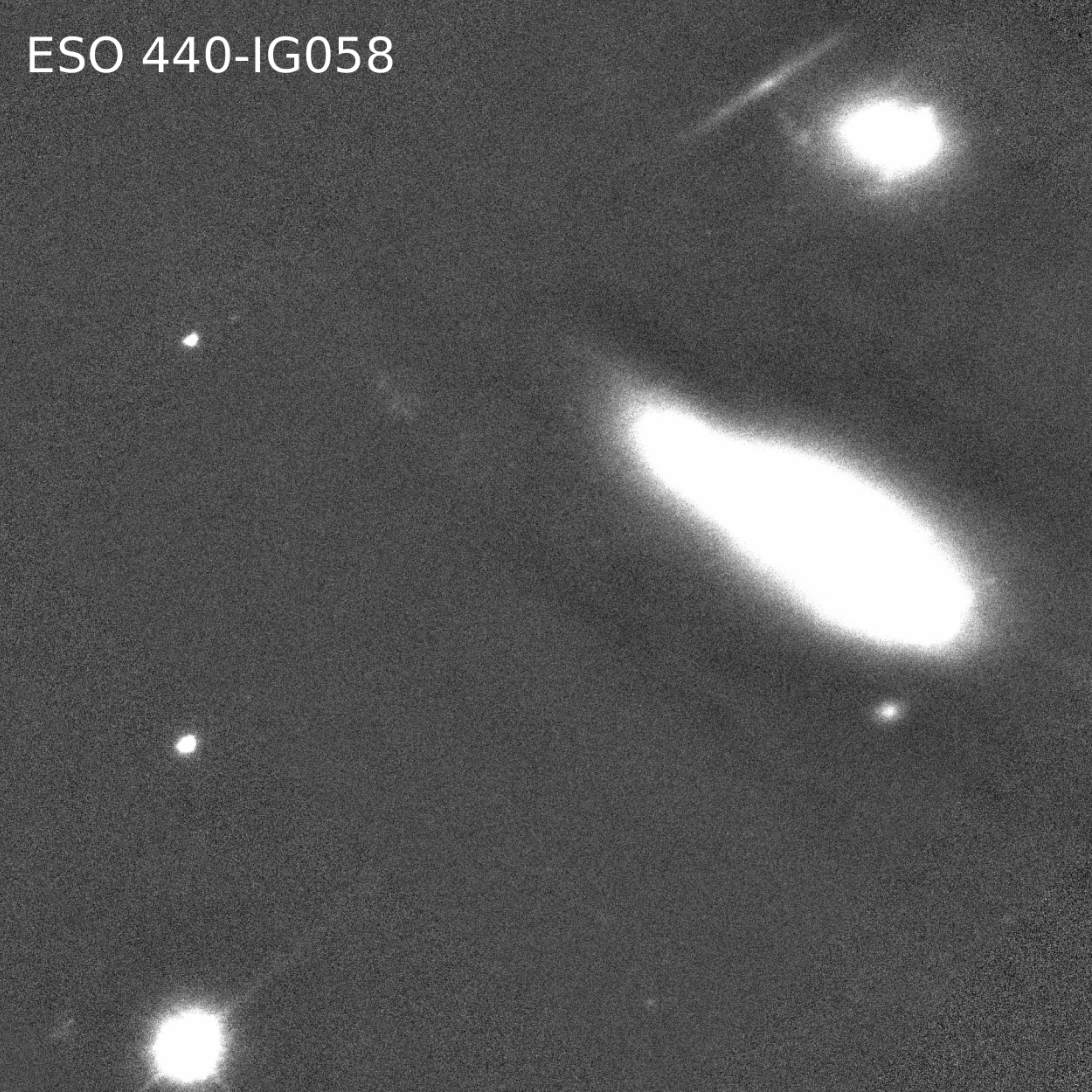}
\includegraphics[width=0.33\linewidth]{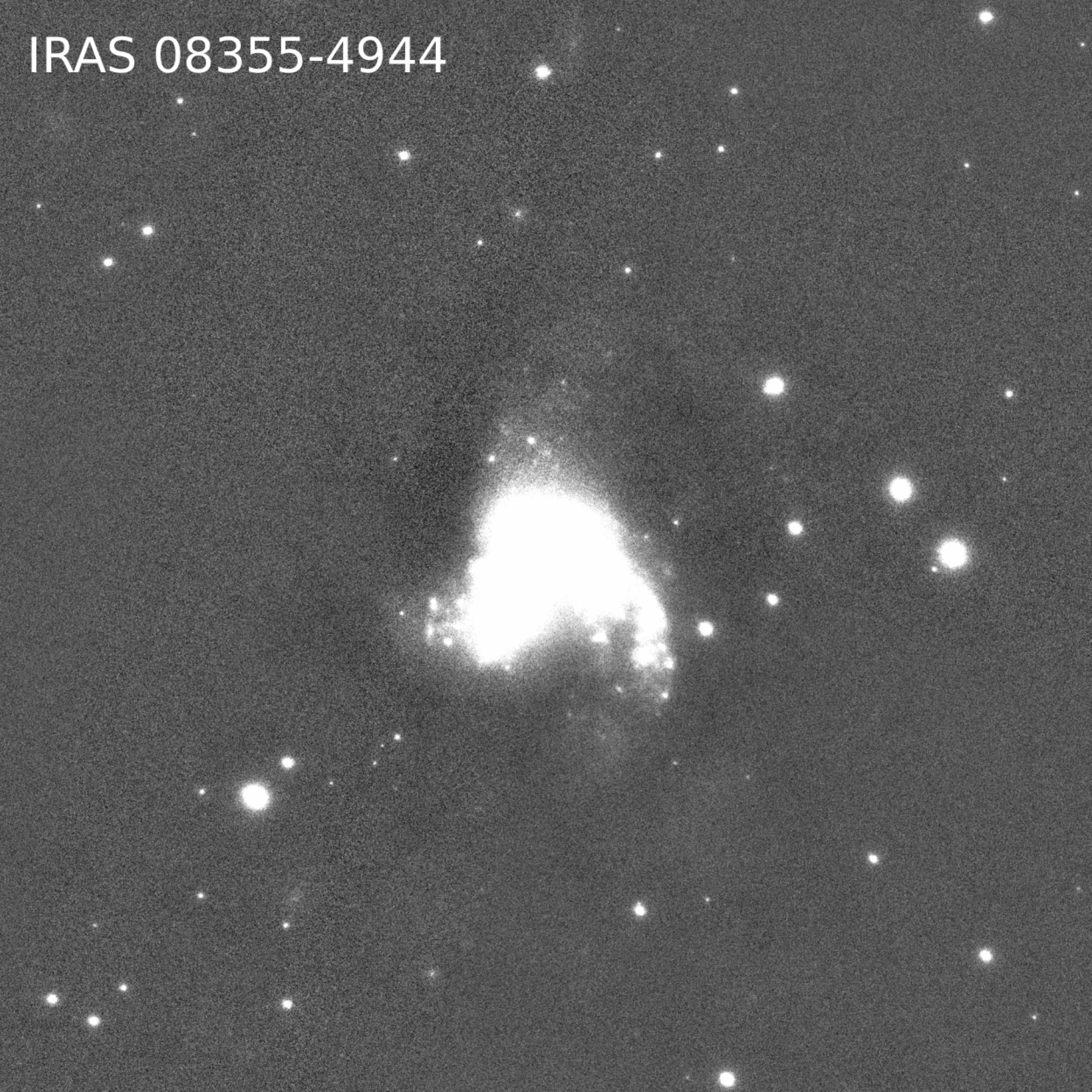}
\includegraphics[width=0.33\linewidth]{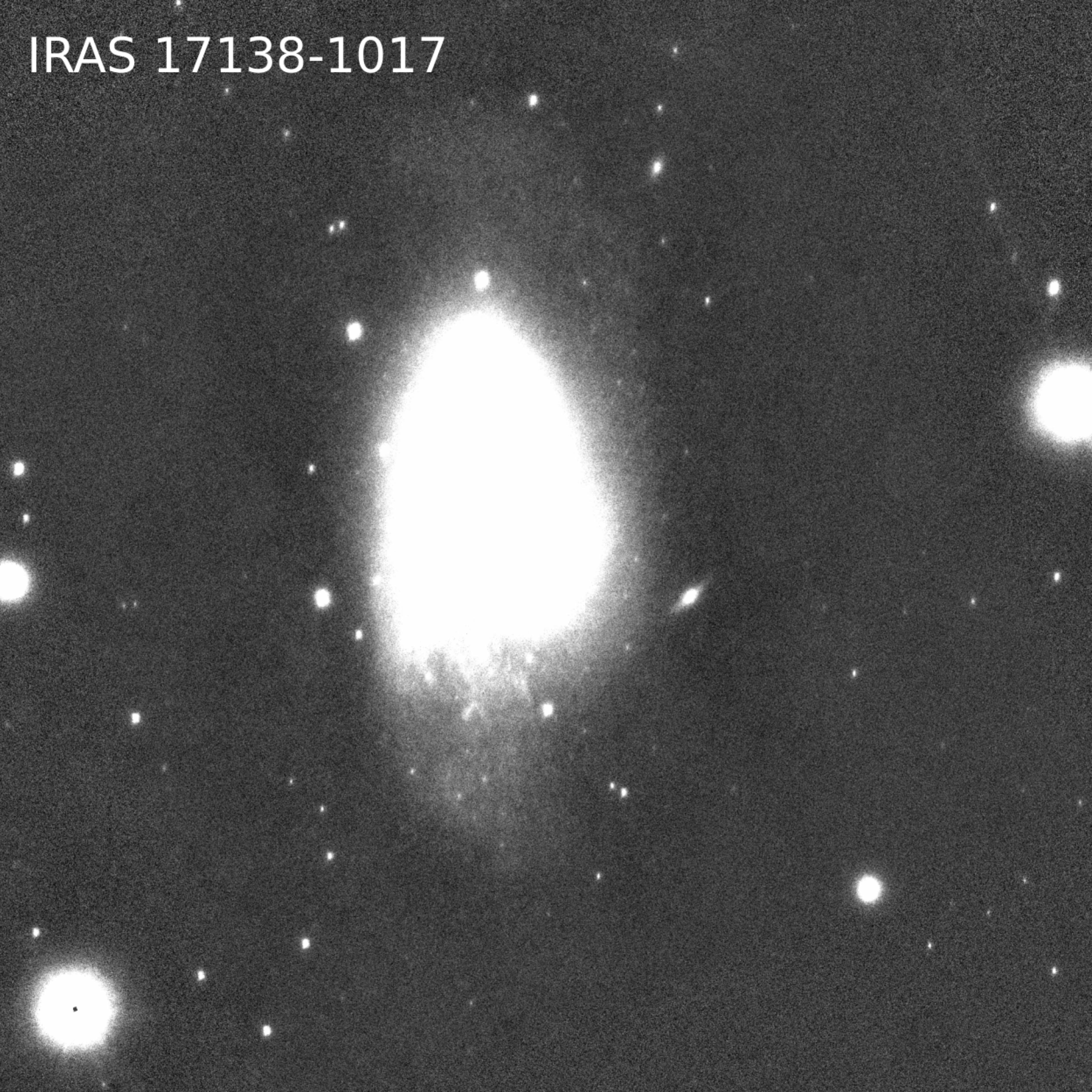}
\includegraphics[width=0.33\linewidth]{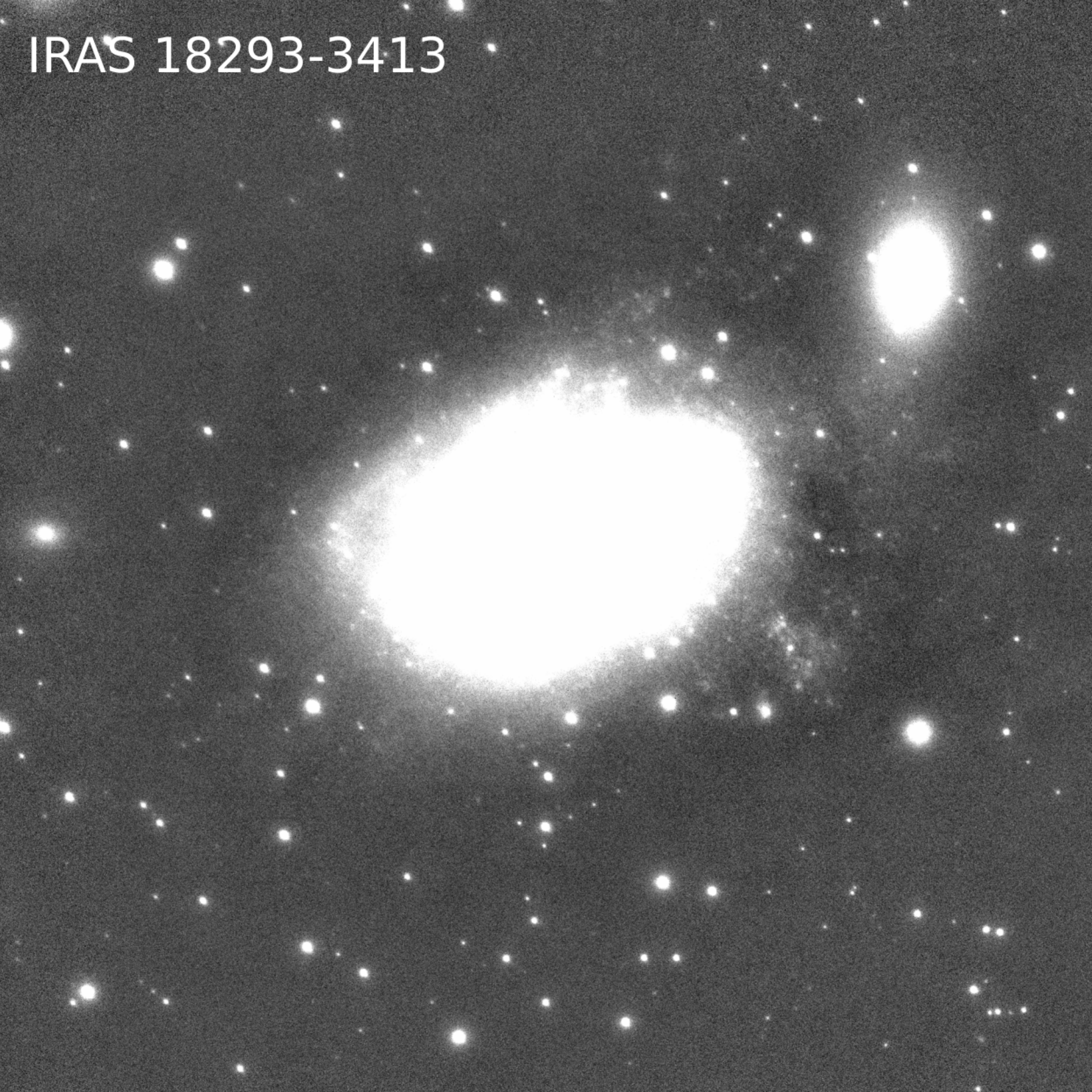}
\includegraphics[width=0.33\linewidth]{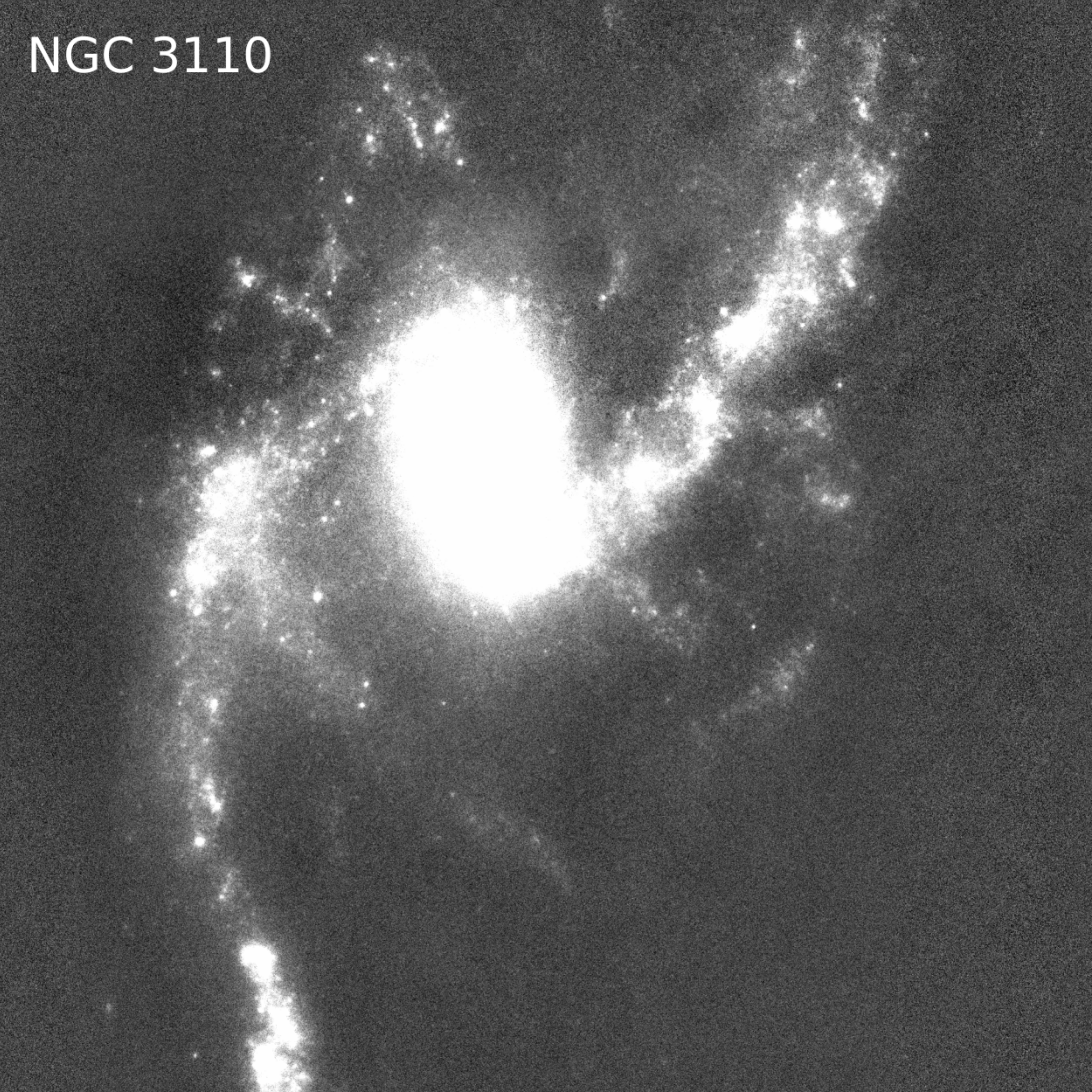}
\includegraphics[width=0.33\linewidth]{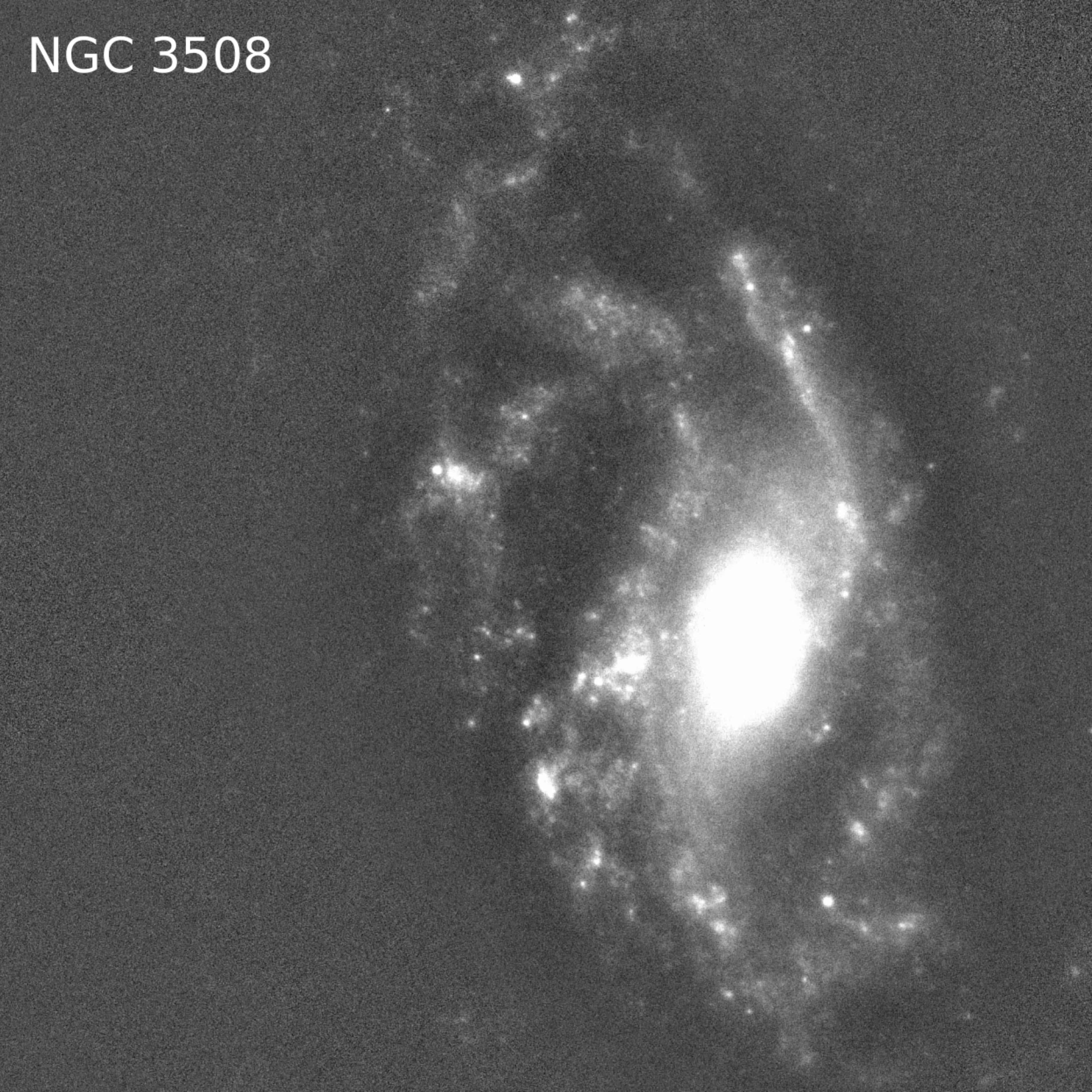}
\includegraphics[width=0.33\linewidth]{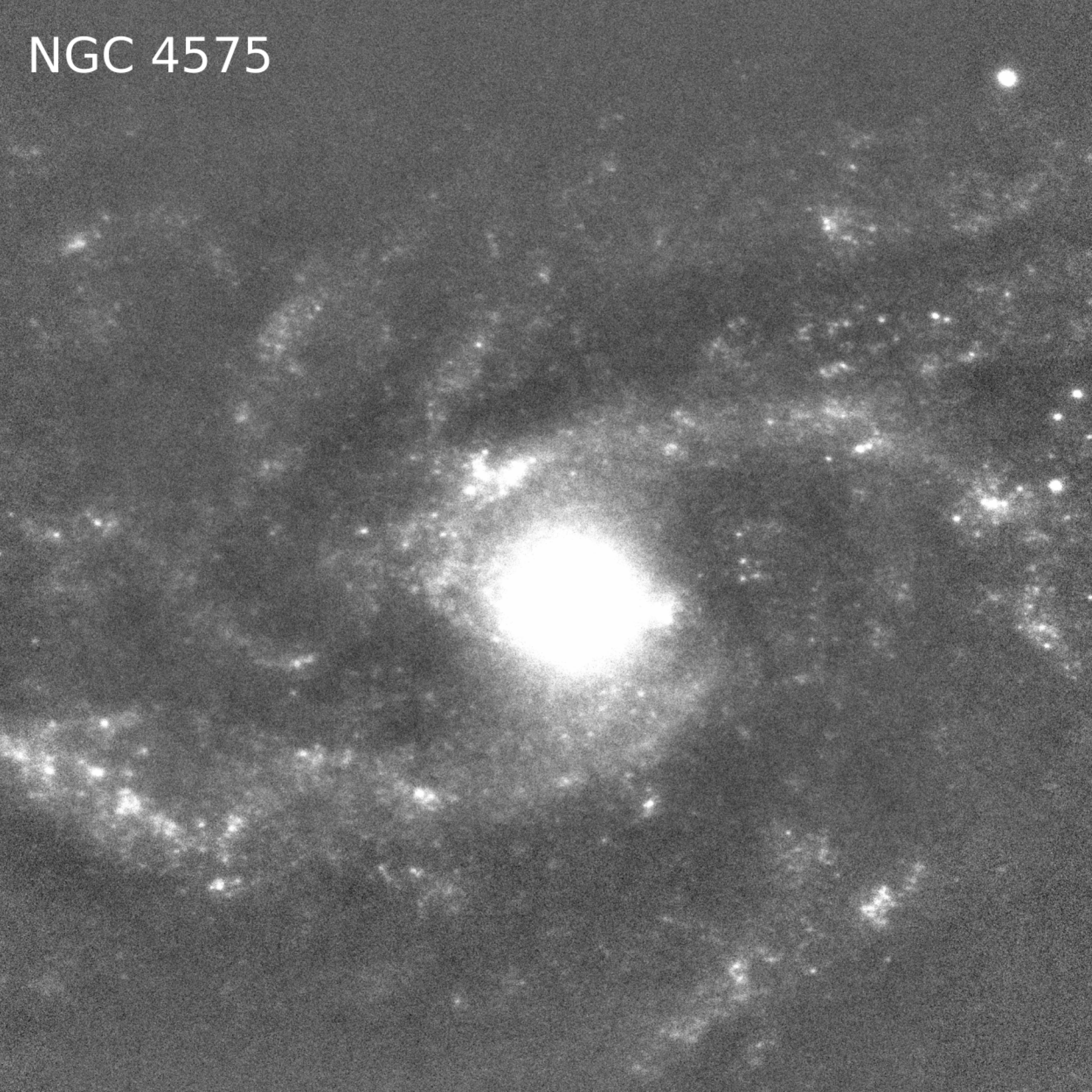}
\caption{Luminous infrared galaxies of the GSAOI/GeMS in 36" $\times$ 36" field of view (except NGC 3110 in 46" $\times$ 46"). North is up, and east is to the left. Due to the pointing of the observations, the target galaxy is not necessarily in the centre of the shown image subsection.}
\label{dataset_examples}
\end{figure*}

    The dataset in this analysis consists of 32 epochs in total from nine different LIRGs, from January 2013 to July 2017 (see Table \ref{data_set_table} and Fig. \ref{dataset_examples}). The sample includes galaxies up to a distance of 115 Mpc, with the LIRGs and the discovered CCSNe presented in the survey paper by \cite{kool2018}. The data were obtained using the Gemini-South telescope with the Gemini South Adaptive Optics Imager (GSAOI, 0.02"/pixel; \citealt{mcgregor2004,eleazar2012}), utilising the Gemini Multi-conjugate Adaptive Optics System (GeMS) in the \textit{K}-band. The system uses five laser guide stars and ideally three natural guide stars to provide an effective and uniform AO correction over the 120" $\times$ GSAOI field of view. The SUNBIRD observations were obtained with the programme IDs GS-2012B-SV-407, GS-2013A-Q-9, GS-2013B-Q-65, GS-2015A-C-2, GS-2015A-Q-6, GS-2015A-Q-7 (PI: Ryder), and GS-2016A-C-1 (PI: Kool). Furthermore, we complemented the analysis with archival data obtained with programme IDs GS-2014A-Q-21 (PI: Lai), GS-2014B-C-1 (PI: Sharp), GS-2015A-C-2 (PI: Sweet), and GS-2016A-C-2 (PI: Sweet) of the sample galaxies with the same \textit{K}-band AO setup and similar image depth. Four LIRGs (ESO 264-G057, ESO 491-G020, MCG +02-20-003, and NGC 1204) were observed only once with GSAOI/GeMS in the SUNBIRD programme and were not included in the analysis due to the lack of a template image of similar quality. Typical on-source exposure times of the dataset were 9 $\times$ 60 s; however, in some epochs a fraction of the frames in the sequence were excluded from the final stacked image of aligned exposures due to poor delivered image quality. 
    
    The flat field correction and sky subtraction steps were carried out with the \texttt{Python} based \texttt{DRAGONS} package \citep{labrie2023}. No separate sky frames were taken for background subtraction and instead the background subtraction was carried out using a sky frame stacked from the individual dithered images. Alignment and co-addition of the reduced sequence images was performed based on the \texttt{IRAF} \citep{tody1986} tasks \texttt{geomap}, \texttt{geotran}, and \texttt{imcombine}. The photometric calibration was based on the Two Micron All Sky Survey (2MASS; \citealt{skrutskie2006}). The sample galaxy NGC 4575 had no 2MASS point sources in the reduced GSAOI/GeMS images and was therefore observed using the Nordic Optical Telescope (NOT; \citealt{NOT_telescope}) with the NOT near-infrared Camera and spectrograph (NOTCam) instrument via the NOT Unbiased Transient Survey 2 (NUTS2) programme. The NOTCam image was reduced using the external \texttt{notcam} package in \texttt{IRAF}, which included flat field correction, sky subtraction, and co-adding individual frames. The 2MASS calibrated NOT image was used to bootstrap magnitudes for several point sources in the GSAOI/GeMS images of NGC 4575.
    The aperture photometry was carried out using the \texttt{Source Extractor} program \citep{bertin1996}.

    Four CCSNe were detected in the dataset: SN 2013if (Type IIP) in IRAS 18293-3413, SN 2015ca (Type IIP) in NGC 3110, SN 2015cb (Type II) in IRAS 17138-1017, and AT 2015cf (unclassified; see Sect. \ref{at_2015cf}) in NGC 3110 \citep{kool2018}. None of the detected transients were observed spectroscopically and the classifications were based on multi-band and multi-epoch photometry. This work did not discover any previously unreported transients in the GSAOI/GeMS dataset.

    We obtained SEDs for seven LIRGs that had not been previously analysed for intrinsic CCSN rates similar to our previous study in Paper I. The LIRGs were: ESO 264-G036, ESO 267-G030, ESO 440-IG058, IRAS 08355-4944, NGC 3110, NGC 3508, and NGC 4575. The SEDs were constructed from a combination of archival photometry and spectra. The photometry was obtained from NASA/IPAC Extragalactic Database (NED) and included data from the Galaxy Evolution Explorer (GALEX, \citealt{galex}), the 2MASS programme, the wide-field Infrared Survey Explorer (WISE, \citealt{wise}), the Infrared Astronomical Satellite (IRAS, \citealt{iras}), the Herschel Space Observatory PACS/SPIRE \citep{pacs}, the Spitzer Space Telescope IRAC/MIPS \citep{irac}, and the Submillimetre Common-User Bolometer Array (SCUBA, \citealt{scuba}). Additional optical photometry by \cite{vaucouleurs1991} and \cite{lauberts1989} was also used. When choosing between multiple values for a given wavelength band, we chose the measurement with the largest aperture to best account for the flux from the whole LIRG. The Spitzer spectra were obtained from the Great Observatories All-sky LIRG Survey (GOALS) project \citep{armus2009} data access archive and were normalised to the photometry based on the $20-30\mu$m flux.

\begin{table*}[!h]
\caption{Summary of the dataset.}
    \centering
    \begin{tabular}{ccccccccc}
    \hline
    \hline
        Galaxy & \textit{N}\tablefootmark{a} & \textit{t}\tablefootmark{b} & SFR\tablefootmark{c} & $t_{\mathrm{SB}}$\tablefootmark{d}& CCSN rate\tablefootmark{e} & $L_{\mathrm{AGN}}$\tablefootmark{f} & \textit{D}\tablefootmark{g} & $A_{\mathrm{Gal},K}$\tablefootmark{h}\\
          &  & (d) & ($M_{\odot}$ yr$^{-1}$) & (Myr) & (SN yr$^{-1}$) & (\%) & (Mpc) & (mag) \\
    \hline
        ESO 264-G036 & 4 & 766 & $76^{+9}_{-13}$ & $34^{+1}_{-2}$ & 0.7$^{+0.1}_{-0.1}$ & 19$^{+5}_{-2}$ & 98.7 & 0.052 \\
        ESO 267-G030 & 4 & 1064 & $22^{+12}_{-6}$ & $25^{+7}_{-4}$ & $0.2^{+0.1}_{-0.1}$ & $29^{+3}_{-3}$ & 95.9 & 0.029 \\
        ESO 440-IG058 & 4 & 1116 & $106^{+7}_{-4}$ & $36^{+1}_{-1}$ & $1.0^{+0.1}_{-0.1}$ & $18^{+5}_{-2}$ & 110.6 & 0.023 \\
        IRAS 08355-4944 & 2 & 356 & $45^{+2}_{-8}$ & $25^{+1}_{-2}$ & $0.6^{+0.1}_{-0.1}$ & $42^{+8}_{-2}$ & 114.9 & 0.250 \\
        IRAS 17138-1017 & 6 & 801 & $69^{+37}_{-4}$ & $28^{+7}_{-1}$ & 0.8$^{+0.1}_{-0.1}$ & 31$^{+20}_{-4}$ & 82.2 & 0.207 \\
        IRAS 18293-3413 & 4 & 772 & $275^{+50}_{-42}$ & $40^{+1}_{-5}$ & $2.3^{+0.4}_{-0.2}$ & $49^{+13}_{-9}$ & 84.1 & 0.043 \\
        NGC 3110 & 4 & 408 & $20^{+6}_{-5}$ & $20^{+6}_{-2}$ & $0.2^{+0.1}_{-0.1}$ & $30^{+3}_{-3}$ & 77.9 & 0.011 \\
        NGC 3508 & 2 & 352 & $8^{+1}_{-1}$ & $37^{+1}_{-2}$ & $0.1^{+0.1}_{-0.1}$ & $16^{+3}_{-2}$ & 61.2 & 0.015 \\
        NGC 4575 & 2 & 353 & $29^{+7}_{-3}$ & $35^{+3}_{-2}$ & $0.2^{+0.1}_{-0.1}$ & $10^{+3}_{-6}$ & 62.6 & 0.037 \\
    \hline
    \end{tabular}
    \label{data_set_table}
    \tablefoot{The luminosity distances are corrected for the influence of the Virgo Cluster and the Great Attractor infall, and assumed $H_0 = 70$ km s$^{-1}$ Mpc$^{-1}$, $\Omega_{\mathrm{M}}=0.3$, $\Omega_{\Lambda}=0.7$. The SFR, CCSN rate, and $L_{\mathrm{AGN}}$ values for IRAS 17138-1017 and IRAS 18293-3413 are adopted from \cite{kankare2021}. \tablefoottext{a}{number of epochs,} \tablefoottext{b}{survey coverage in days,} \tablefoottext{c}{estimated SFR value,} \tablefoottext{d}{starburst age,} \tablefoottext{e}{intrinsic rate of CCSNe,} \tablefoottext{f}{AGN contribution to the luminosity,} \tablefoottext{g}{luminosity distance,} \tablefoottext{h}{Galactic line-of-sight extinction.}}
\end{table*}

\section{Method}
\label{method}

    Our method can be separated into four distinct parts: (1) automated constraining of optimal template subtraction parameters, (2) determining limiting magnitudes of the images via artificial CCSN injection, (3) deriving CCSN detection probabilities with a Monte Carlo method, and (4) yielding the undetected fraction by comparing the detection probabilities with the intrinsic CCSN rates derived via SED fitting for the dataset (\ref{intrinsic_rates}). Brief descriptions of these steps are presented in the following subsections. For a more detailed description, see Paper I.

\subsection{Automatic image subtraction}
\label{automatic_image_subtraction}

    We used the method described by \cite{james2006} to define the pixel regions covered by each sample LIRG in the GSAOI/GeMS images (hereafter galaxy maps). First, a median stacked image was created from the reduced and aligned images of a LIRG. The background level was subtracted from the median image. This level was a median value of a selected isolated area from the image, far from the LIRG and other field sources. Lastly, the pixel values were summed from the lowest to the highest. The ranked pixels before the sum reaches zero are considered to belong to the background, and the remaining to the LIRG or other field sources (e.g. foreground stars or unrelated galaxies). These other sources were excluded from the galaxy maps manually. The selection of the background sample area has some effect on the galaxy maps; however, the effect is limited to the faint outer edges of the LIRGs with a minimal contribution to our final results.

    Template subtraction is a standard method used by SN search programs. The template subtraction in our study was performed with a modified version of the ISIS2.2 package \citep{alard1998,alard2000}. In the process, two images are aligned, the image with narrower PSF is convolved to match the other image, and their flux levels are matched. Subsequently, one of the images is subtracted from the other pixel by pixel. Sources that remain constant between the image epochs (e.g. galaxies and stars) should disappear, and transient sources (e.g. CCSNe) should remain, given that the transient is not present in one of the images, or has sufficiently evolved between them. In practice, the subtraction process may leave residuals in the resulting image, particularly to the brightest regions in the images.  

    The template subtraction process is dependent on some of the chosen parameters (e.g. \citealt{melinder2012}). With an aim to reduce the subjective human element, and to standardise the subtraction quality within the dataset, we developed an algorithm to semi-automatically perform the template subtraction as described in detail in Paper I. The user provides the code sets of initial guesses and permitted ranges for the parameters, which the code optimises. This algorithm was used to determine the sets of optimal template subtraction parameters for each image in the GSAOI/GeMS dataset to be used in the later steps. The quality of the subtraction was defined as the variance of the pixel values over the subtracted image in the galaxy map area, excluding the areas (stamps) used to derive the convolution kernel for the image subtraction process.

    The stamp regions that were used for calculating the convolution kernel were not used in the analysis of the resulting subtracted images, since a possible faint transient in those areas would likely disappear in the process. Therefore, the subtraction process needs to be repeated at least twice to cover the whole image. In practice, multiple subtractions with different stamp combinations were performed to ensure that each cycle had sufficient coverage to derive the image convolution for the process.

\subsection{Limiting magnitudes}
\label{limiting_magnitudes}

    Noise in the template subtracted images and residuals from the image subtraction process can have a major impact on the ability to recover transients from the resulting images. These effects are more prominent in regions with higher flux and stronger gradients (e.g. the LIRG nuclei or bright super star clusters). Thus, the recoverability of objects can greatly vary across the image, and spatially resolved estimates for the limiting magnitudes are required.

    The code for this process was created in Paper I, and the GSAOI/GeMS dataset was analysed with the same approach. The method was to create artificial SNe with the \texttt{IRAF} routine \texttt{mkobjects} at different brightness levels in the images, carry out template subtraction process, and attempt to recover the simulated sources. The FWHM of the artificial SNe (Moffat profile parameter 2.5) was matched with the FWHM of the narrow core of the AO PSF of point sources in the original images. 

    The positions of the artificial SNe were determined by dividing the image into a grid with a spacing equal to the FWHM of that epoch. An artificial SN was created in the grid location if at least one galaxy map pixel was within it. Creating all these artificial SNe simultaneously would result in blending and render detecting them individually impossible. Therefore, the artificial SNe were created in groups, where they would be separated by at least seven times the FWHM of the image. This was found to be sufficient to prevent any notable overlap effect. 

    The following iterative process was used to determine the limiting magnitudes of an epoch. Artificial SNe with an apparent brightness of 16 mag were created on the image for which the image subtraction process was carried out. The artificial SNe were detected from the subtracted image with \texttt{Source Extractor}. The process was repeated with artificial SNe that were 0.2 mag fainter, and continued until no detections were made or the process had repeated 40 times. If there were areas where the starting brightness of 16 mag was not sufficient, these were re-analysed with a starting brightness of 14 mag. The threshold for a detection was set as a minimum of 9 pixels which are at least 5$\sigma$ above the background level. Areas within 10 $\times$ FWHM of any previously detected real SNe in the dataset were excluded from the analysis. The minimum pixel count was chosen to exclude artefacts that cover only a few pixels in the subtracted images and are clearly not real. The conservative 5$\sigma$ threshold was chosen by simulating sources at varying thresholds and choosing the limit at which the simulated sources were still readily detected by eye from the often noisy subtracted central regions of the sample galaxies.

    Residuals from the subtraction process can cause false positives. This is further complicated when an artificial SN is created on or near an artefact. It becomes difficult to ascertain at which point the combined artificial SN and residual feature should no longer be counted as a detection. In these cases, the region was conservatively set as an undetectable region. This effect was mitigated by multiple subtractions with different sets of image subtraction stamps. 
    The process described produces multiple sets of parameters for each epoch. The iterative process was performed for each of these. This resulted in each pixel in the galaxy map to have at least one limiting magnitude value. In cases where two or more values existed, the faintest magnitude limit was adopted. 

\subsection{Monte Carlo method}
\label{monte_carlo_method}

    Core-collapse supernovae remain undetected in surveys due to a combination of factors: low cadence, insufficient imaging depth and resolution, poor image subtraction performance in nuclear regions, or high host extinction. Disentangling the effect of host extinction is not straightforward; however, we aimed to constrain this by simulating the SUNBIRD GSAOI/GeMS survey with varying degrees of extreme dust obscuration using a Monte Carlo method. Subsequently, we matched the expected CCSN detections of the simulation to the real detections from the dataset to derive the missing fraction and the undetectable fraction. The process was developed in Paper I, which also contains a more detailed description of the process.

    The components of the Monte Carlo simulation were the survey cadence, detection thresholds for the faintest detectable CCSNe in each pixel in each epoch of the dataset, CCSN light curve templates and their variability ranges, host extinction model, the relative rates of CCSN subtypes, and the estimated intrinsic CCSN rates for the sample galaxies.

    The time resolution of the simulation was one day, spatial resolution one pixel, and each run simulated $10^7$ SNe. The following procedure was used. First, a CCSN subtype was chosen. Then an explosion date was randomly chosen within the typical detection window (i.e. control time; see Sect. \ref{control_time}). Then an explosion location was chosen from the galaxy map randomly with the probability directly proportional to the host galaxy \textit{K}-band flux. Next, the light curve template was generated, and a host extinction, a galactic extinction, and a distance modulus were applied. The resulting apparent magnitude was then compared against the limiting magnitude of the explosion site pixel in the first epoch where the SN would be observable. If the SN was brighter, it was counted as a detection.

\subsubsection{Template light curves}
\label{template_light_curves}

    The subtypes of CCSNe considered in this work are: Ib, Ic, IIb, II, IIn, and SN 1987A-like. Similar to Paper I, the relative rates of these  CCSN subtypes were adopted from the observed volume limited results of the Lick Observatory Supernova Search (LOSS, \citealt{Li2000}) with the CCSN classifications revised by \cite{shivers2017}. This work adopts the CCSN peak magnitudes and their relative rates with small modifications. We combined the subtypes Type II-P and II-L into one population of Type II SNe. Similarly, the peculiar subtypes (e.g. Type IIb-peculiar) were combined with their respective non-peculiar subtypes. The narrow and broad line subtypes other than the Type IIn were also combined with their respective normal subtypes.

    We created CCSN subtype light curve templates to be used in the Monte Carlo simulation in Paper I and briefly explain the process here.
    We chose a high quality \textit{K}-band light curve of a CCSN for each subtype, that is a good representative of that subtype. The peak magnitudes of the templates are based on statistics of the optical LOSS sample revised by \cite{shivers2017} from which the \textit{K}-band peak magnitudes were estimated based on \textit{R-K} colour ranges adopted for different CCSN subtypes in Paper I.
    
\begin{figure}[hbt]
    \begin{center}
        \includegraphics[width=0.5\textwidth]{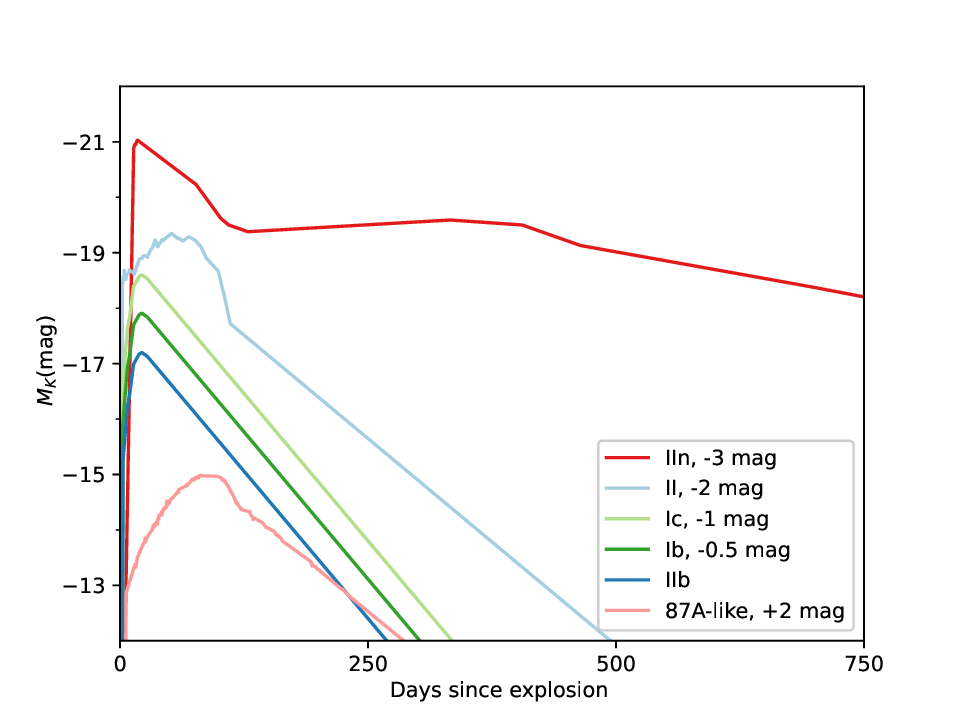}
        \end{center}
    \caption{Template \textit{K}-band light curves of CCSN subtypes used in the Monte Carlo simulation. The light curves have been shifted vertically for clarity. The magnitudes are in the Vega system.}
    \label{light_curves_figure}
\end{figure}

    To emulate the time-scale variability of CCSNe, we applied a time factor to the Type II and the stripped envelope templates. Similar to Paper I, we scaled the plateau phase of the template light curve to match the mean of the \cite{anderson2014} sample and set the time factor range to match the variance of that sample. The time factor is then applied to the template plateau phase when the templates are generated in the simulation. A similar approach was taken with the stripped envelope template, except instead of the plateau length, the difference in magnitude between the peak and 15 days after the peak ($m_{\Delta15}$, \citealt{delta_15}) was used to scale the photospheric phase of the light curve. The shortest possible time factor was restricted to match the fastest evolving SN in the colour comparison sample to exclude unphysically rapidly evolving templates. The time factor was applied to the first 35 days of the template to avoid unrealistic late-phase evolution. The adopted parameters for the CCSN subtype templates and ranges for their light curve variability are listed in table 1 of Paper I. The most likely templates of each CCSN subtype are presented in Fig. \ref{light_curves_figure}.

\subsubsection{Host extinction}
\label{host_extinction}

    Galactic extinction values were obtained from NED, which are based on the dust map calibration of \cite{schlafly2011}. These are presented in Table \ref{data_set_table}.
    \cite{kankare2021} reported host extinction estimates of 19 CCSNe discovered via optical or IR methods in the central regions of LIRGs with mean and median values of $A_V = 3.6$ and 2.1 mag, respectively, which included the highest measured host extinction for an IR detected CCSN with an $A_{V} = 16$ mag \citep{kankare2008}. Furthermore, central regions of LIRGs can have extinctions of several hundreds of $A_V $, with some up to $A_V \gg 1000$ mag \citep{torres2021}. 
    For comparison, \cite{pessi2023} measured host extinctions for 71 CCSNe in normal galaxies from the All-Sky Automated Survey for Supernovae (ASAS-SN, \citealt{shappee2014,kochanek2017}), resulting in mean values of $A_{V} = 0.23 \pm0.18$ and $0.30 \pm0.16$ mag for Type II and stripped envelope subtypes, respectively. It is clear that especially the optical observations are extremely biased towards lower extinctions and can only probe the surface layers of the LIRGs. We address this by dividing the CCSNe into two populations by their host extinctions, a surface component and a deep component. 
    
    The CCSN sample by \cite{kankare2021} was used to describe the surface extinction component. The data were fit with an exponential function $P(A_V)=0.881e^{-0.084A_V}$ with a median value of $A_V = 7.7$ mag in Paper I. This extinction model is evaluated up to $A_{V} = 40$ mag, which roughly corresponds to the most highly extinguished CCSN that could reasonably be detected in the dataset. The simulated CCSNe were randomly assigned host extinctions from this distribution. 
    
    The deep component describes the population of CCSNe that explode deep in the dusty clouds in the central regions of the LIRGs and is completely undetectable with optical and IR surveys. We refer to this as the missing fraction. Each run of the Monte Carlo simulation had a set probability for any simulated CCSN to be assigned in the missing fraction, which was varied between subsequent runs from 0 to 99 \%.

\subsubsection{Control time}
\label{control_time}

    Transient surveys with a more limited cadence may have SNe explode and fade below the detection limit before the next survey epoch. This results in the effective survey time being shorter than the survey duration bound by the first and the last epochs, and should be accounted for when calculating the expected SN rates. Likewise, simulating SN explosions on epochs where the time to the next survey observations is too long introduces a bias towards a lower missing fraction. Therefore, in our Monte Carlo method, the simulated CCSNe can only explode during time windows where the time to the next epoch is within the control time (CT) of that CCSN subtype for that LIRG, including before the first epoch. CT is a measure of how long a SN could be typically detected, and is dependent on the SN subtype and the survey depth (see e.g. \citealt{cappellaro1993}). 

    Our method of calculating the CT for a given CCSN subtype and epoch was the following: $10^5$ light curves were randomly generated based on the parameters introduced in Sect. \ref{template_light_curves}. The distance modulus and random host extinction were applied as described in Sect. \ref{host_extinction}. The CT was calculated as the mean time the light curves remain above a magnitude limit. The magnitude limit was a weighted average of the magnitude limits of an epoch (Sect. \ref{limiting_magnitudes}), where the weight was proportional to the flux of that pixel relative to the total flux of the galaxy. This was repeated for all CCSN subtypes and for all epochs. The CTs for the sample galaxies are listed in Tables \ref{ct_ESO_264-G036}-\ref{ct_ngc_4575} in the Appendix. The median CT of the fastest evolving CCSN template IIb was 109 days.

    Total control time (TCT) is a measure of the effective survey time for a galaxy. It is calculated as the sum of times between epochs, except when the CT corresponding to the latter epoch is shorter than the time between epochs. The TCT is used to calculate the underlying CCSN count for each of the CCSN subtypes.

\subsubsection{Intrinsic CCSN rates}
\label{intrinsic_rates}

    We estimated intrinsic CCSN rates for the LIRGs in the dataset with an adapted version of the SATMC Monte Carlo code \citep{SATMC,efstathiou2022}. SATMC method fits several components to the SED of the sample LIRGs to determine their relative contributions. The components were as follows: starburst \citep{SED_starburst1,SED_starburst2}, AGN \citep{SED_AGN1,SED_AGN2}, and a spheroidal galaxy \citep{SED_spheroid} or a disc galaxy \citep{efstathiou2022}. The SED of the starburst component determines the age of the starburst episode, which was allowed to range within 5 to 50 Myr. The starburst model incorporates the stellar population synthesis model of \citeauthor{bruzual1993} (\citeyear{bruzual1993,bruzual2003}) which combined with the luminosity can estimate the mass of stars that formed in the starburst episode and the resulting CCSN rate based on the age of the starburst and its $e$-folding time, which was allowed to range within 1 to 40 Myr. The method adopts the Salpeter IMF for both the starburst and spheroidal components. The starburst model assumes solar metallicity, whereas the spheroidal component assumes that the metallicity is 40\% of solar.  The same approach was also used in Paper I.

\begin{figure*}
\centering
\includegraphics[width=0.33\linewidth]{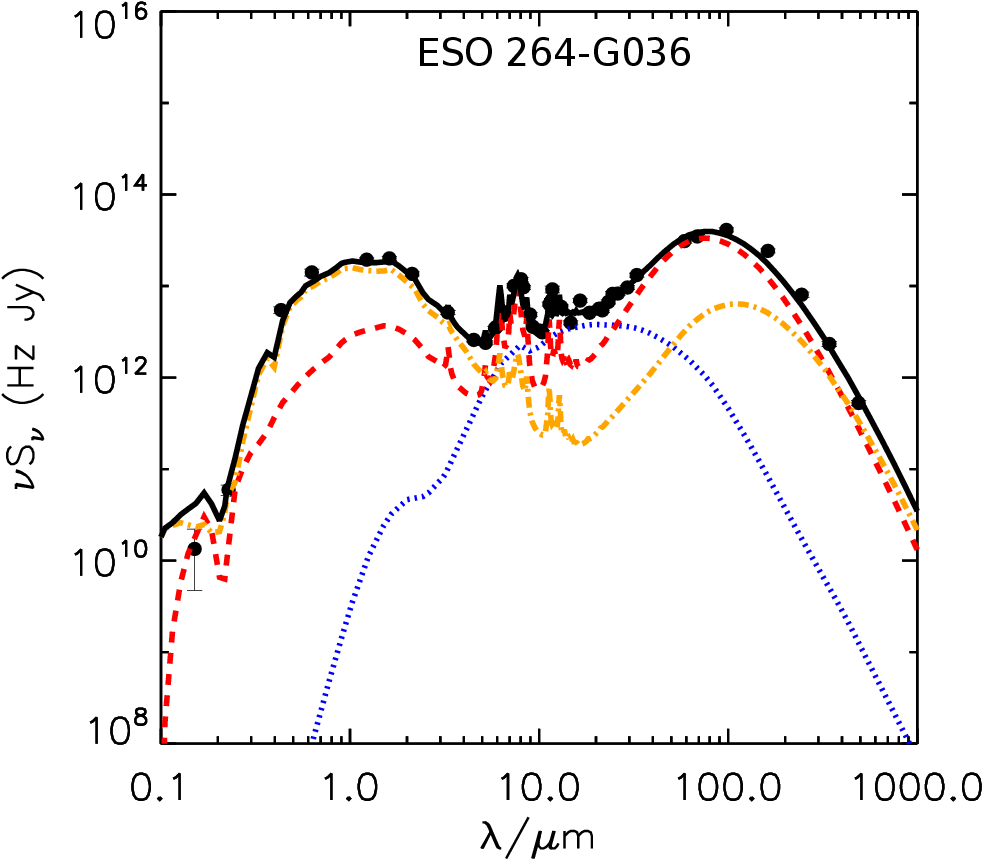}
\includegraphics[width=0.33\linewidth]{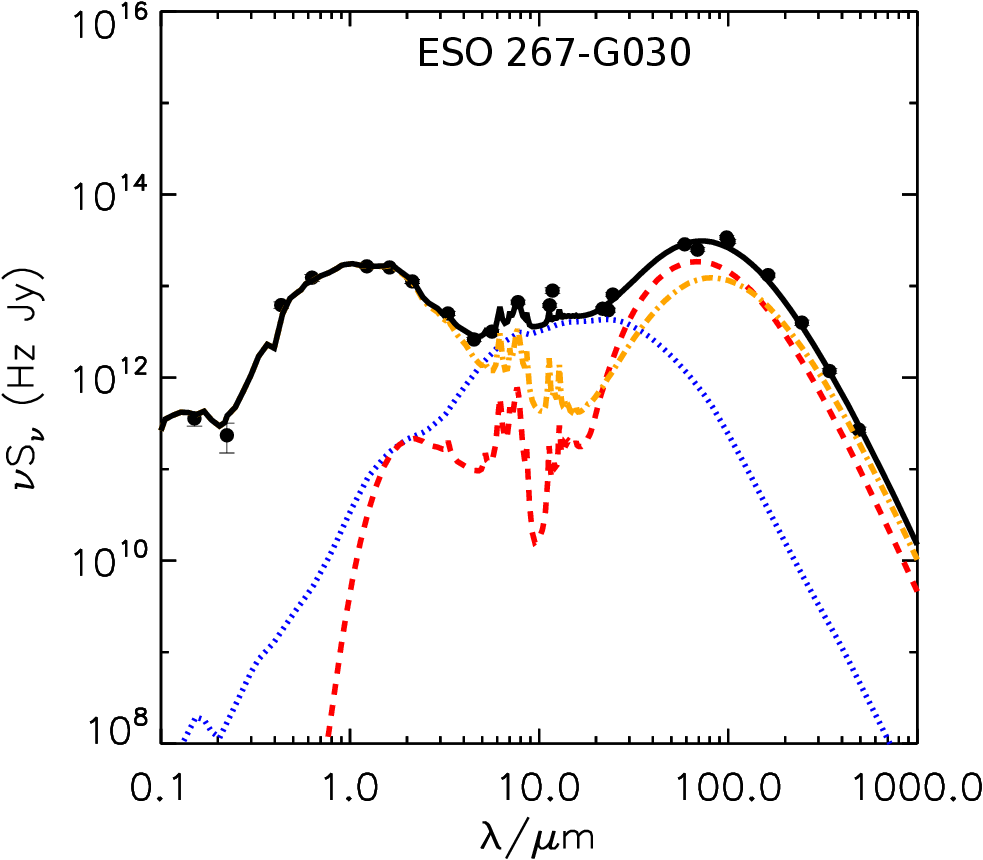}
\includegraphics[width=0.33\linewidth]{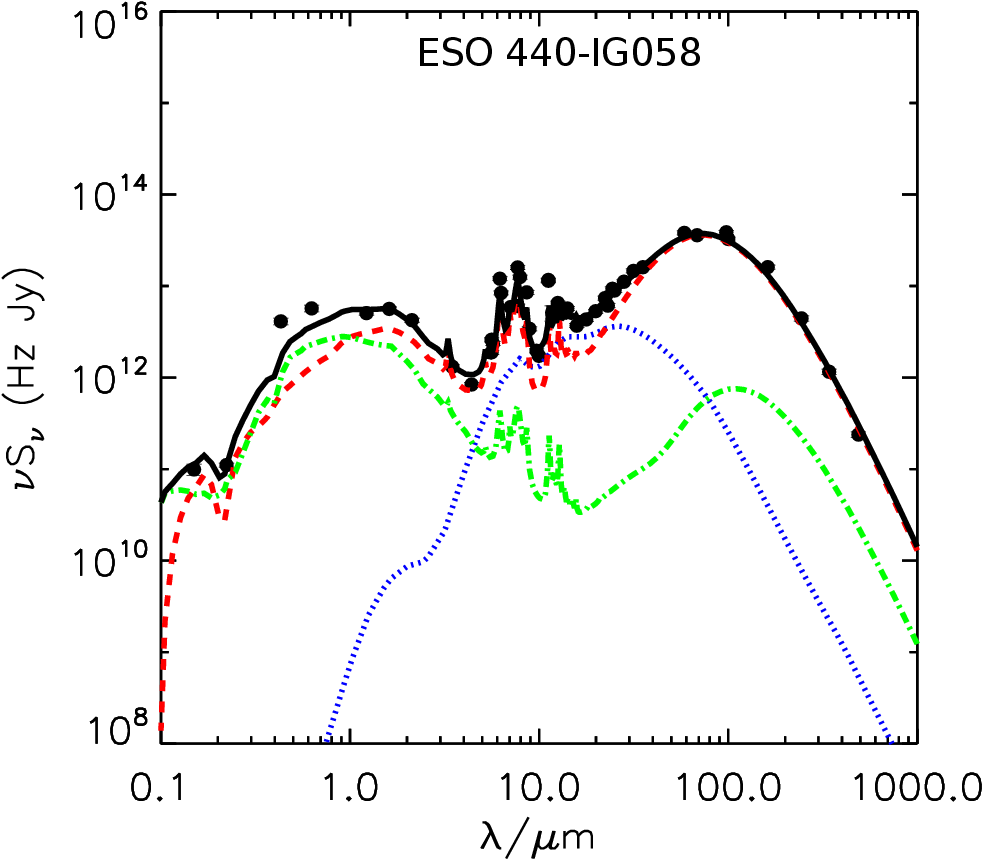}
\includegraphics[width=0.33\linewidth]{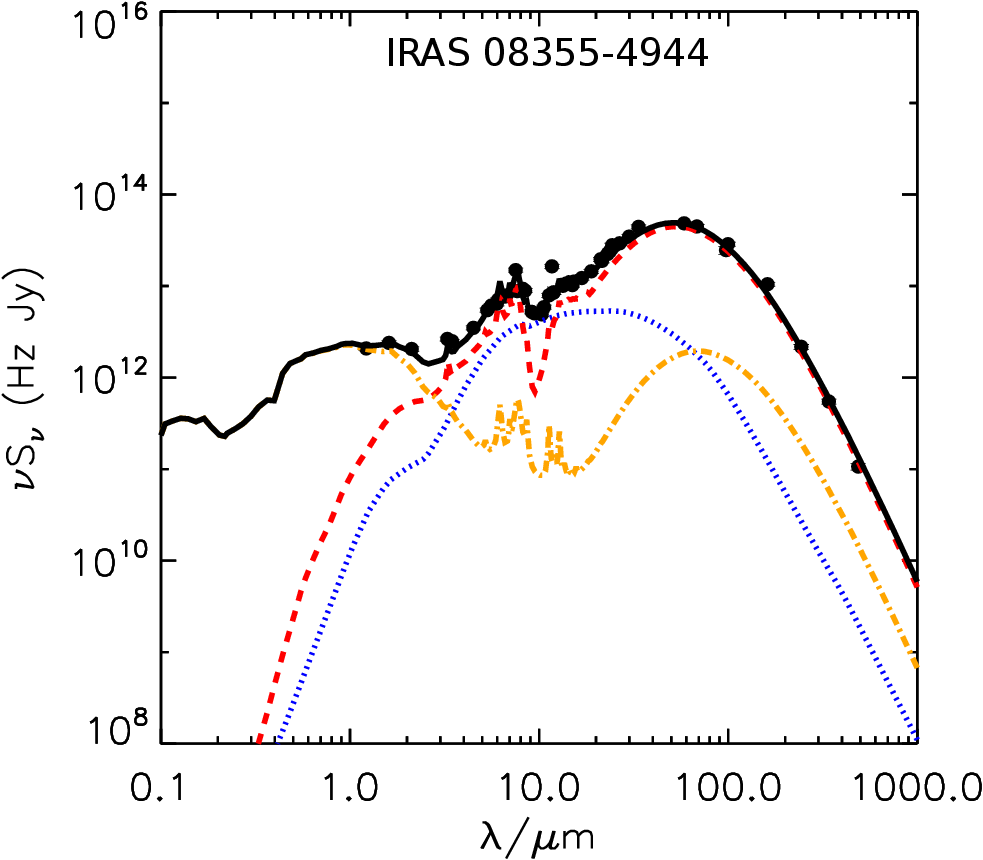}
\includegraphics[width=0.33\linewidth]{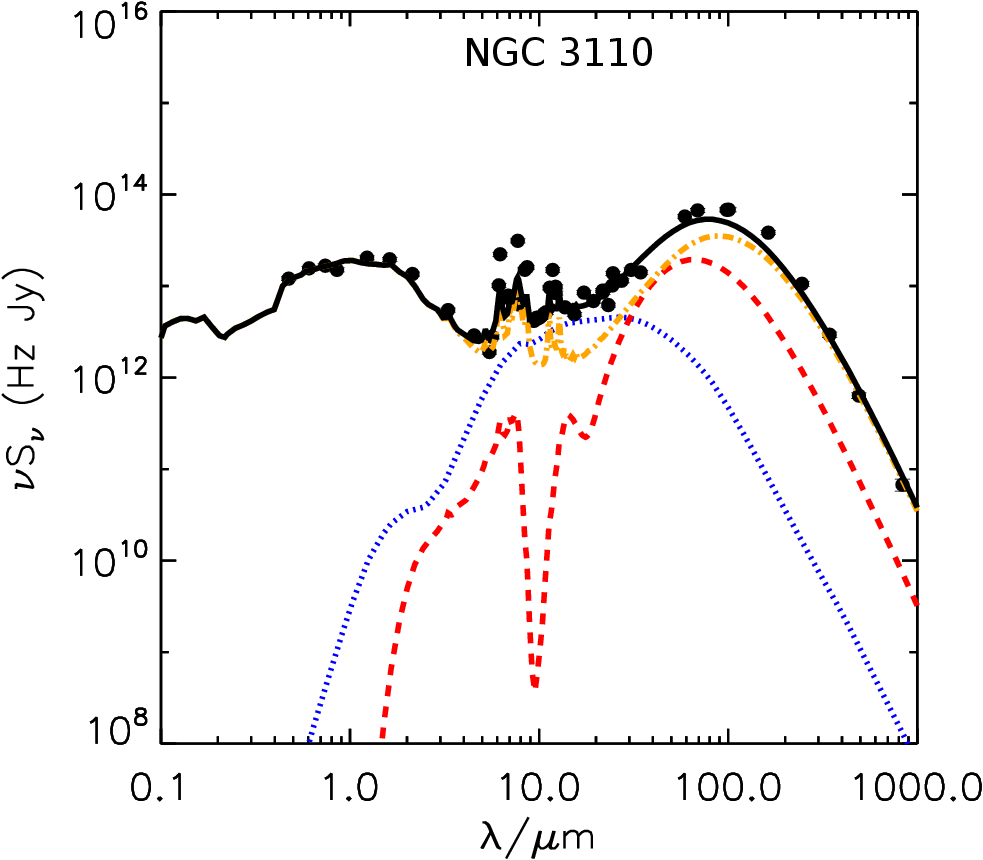}
\includegraphics[width=0.33\linewidth]{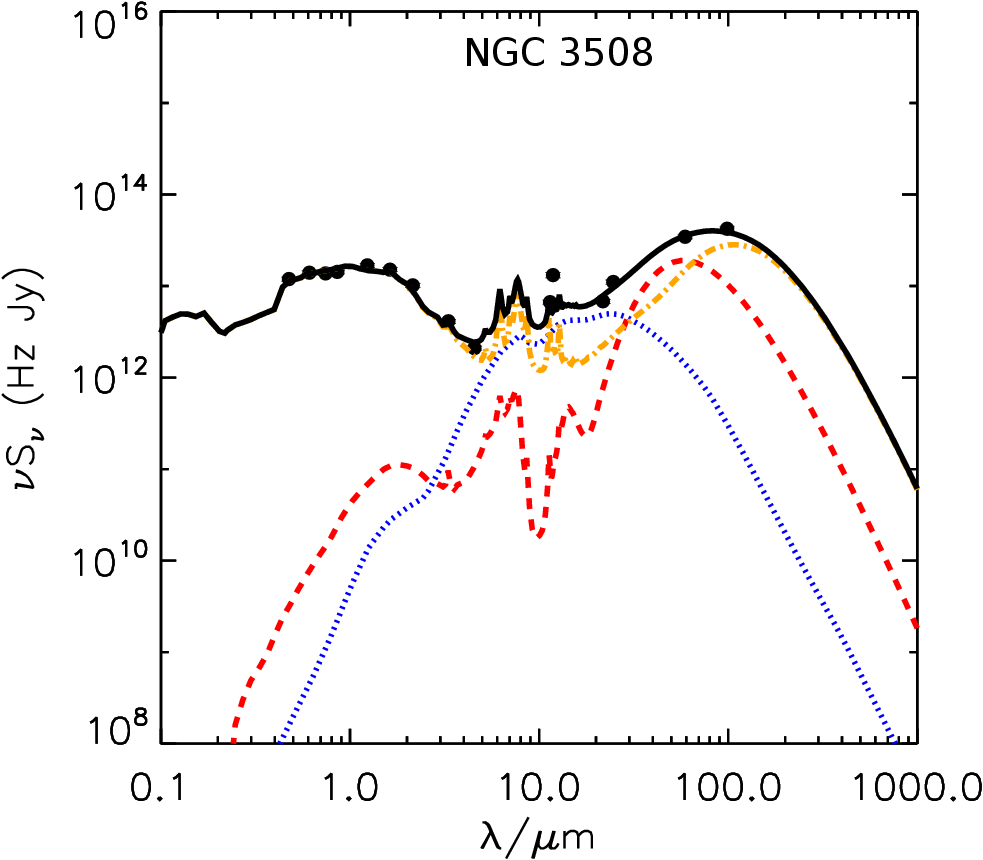}
\includegraphics[width=0.33\linewidth]{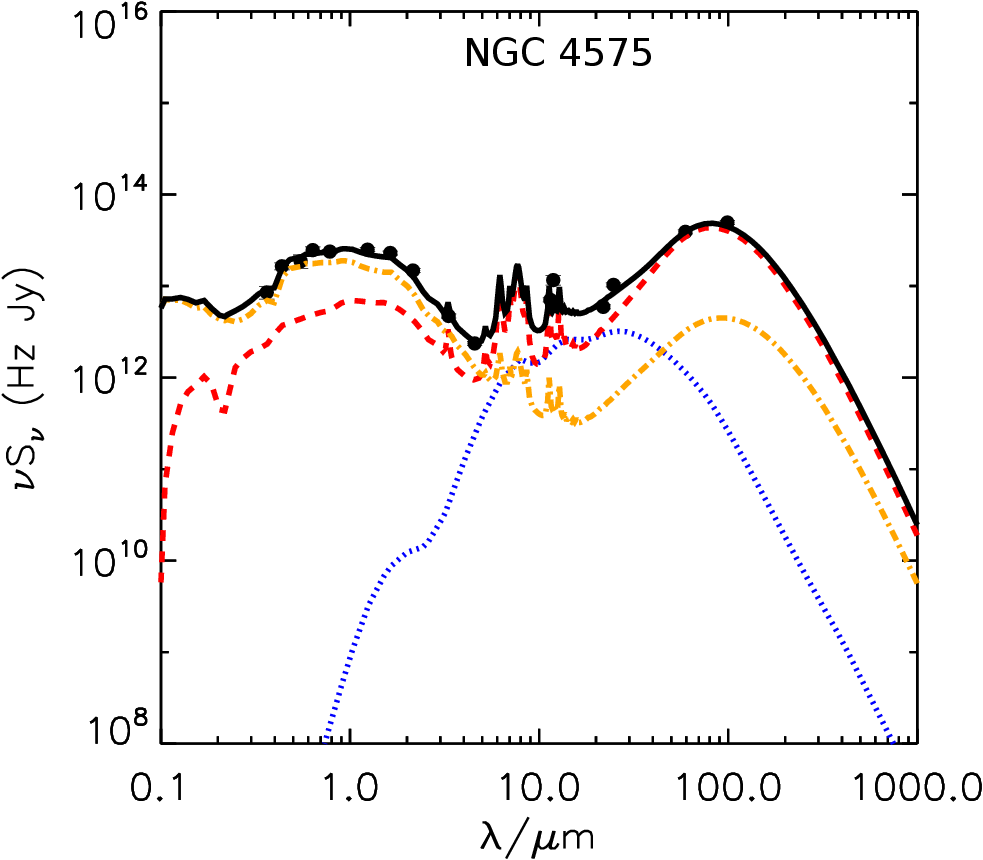}
\caption{Results of the SED modelling (solid black) of seven LIRGs of the GSAOI/GeMS dataset produced in this work. The model components are: either a spheroidal (dot-dashed orange) or a disc (dot-dashed green) galaxy, starburst contribution (dashed red), and an AGN (dotted blue).}
\label{sed_examples}
\end{figure*}

    In this work, we derived the inferred intrinsic CCSN rates for ESO 264-G036, ESO 267-G030, ESO 440-IG058, IRAS 08355-4944, NGC 3110, NGC 3508, and NGC 4575. The intrinsic CCSN rates for IRAS 17138-1017, and IRAS 18293-3413 were obtained from \cite{kankare2021} and derived with the identical method. The estimated intrinsic CCSN rates are summarised in Table \ref{data_set_table}, and the SED fits are shown in Fig. \ref{sed_examples}.

\subsubsection{Error estimation}

    In order to estimate the errors in the Monte Carlo method, we ran the code with various parameters offset by their respective errors. The parameters included and their respective errors were the following. Limiting magnitude determination used a 0.2 mag step size, resulting in an error of $\pm 0.2$ mag. To estimate the uncertainty of the photometric calibration, we measured the standard deviation of the photometric zero point corrections from epochs with the same total exposure time, and conservatively chose the largest one, resulting in an error of $\pm 0.335$ mag. This error estimation reflects the effects of varying AO correction effectiveness on the Strehl ratio. The increased flux in the unresolved PSF wings caused by less effective AO correction results in uncertainty in the photometric calibration. For the colour correction $M_R - M_K$ in the CCSN peak magnitude distributions, we chose the largest standard deviation within a CCSN subtype. This was amongst H-poor templates, resulting in an error of $\pm 0.26$ mag (Paper I). Finally, the error in distance modulus was $\pm 0.15$ mag for all the sample LIRGs, reported by NED. Thus, the resulting total error as the square root of the sum of squares is $\pm 0.49$ mag. Additionally, we applied Poissonian confidence limits by \cite{gehrels1986} to the inferred small number of intrinsic CCSNe in the simulation to address the small number statistics, which dominates the uncertainty in the resulting undetectable fractions of CCSNe in LIRGs.

\subsection{The nature of AT 2015cf in NGC 3110}
\label{at_2015cf}

The GSAOI/GeMS follow-up dataset of AT 2015cf reported by \cite{kool2018} is limited to two epochs of \textit{K}-band detections at $m_{K} = 20.97 \pm 0.13$ and $21.4 \pm 0.2$ mag on JD 2457092.6 and JD 2457172.4, respectively, and two \textit{H} and \textit{J}-band limits of $m_{H} > 22.7$~mag and $m_{J} > 22.8$ mag on JD 2457174.4. The two epochs or detections confirm that the transient is real; however, based on such a small amount of data the nature of the event remained ambiguous. The \textit{K}-band data points suggest a decline rate of $\gamma_{K} = 0.54 \pm 0.30$~mag~(100~d)$^{-1}$, which is slower than the theoretical $0.98$~mag~(100~d)$^{-1}$ decline rate of the $^{56}$Co to $^{56}$Fe decay tail assuming a complete gamma-ray trapping where $^{56}$Co is a decay product of $^{56}$Ni synthesised in a SN explosion. If AT~2015cf is a CCSN, it is obscured by dust, intrinsically faint, and/or observed at late phases, based on the faint \textit{K}-band detections and deep \textit{H} and \textit{J}-band limits. Unfortunately, there are no constraining pre-explosion limits for the event; thus it is plausible that the transient was discovered at a late-time epoch. While there are CCSNe discovered in the central regions of LIRGs with high derived host galaxy line-of-sight extinctions (e.g. SN~2008cs with $A_{V}^{\mathrm{host}} = 16$~mag at a projected distance of 1.5~kpc; \citealt{kankare2008}) the apparent location of AT~2015cf in one of the spiral arms of NGC~3110 is quite removed at a projected distance of 3.5~kpc. If not a CCSN, the most likely alternatives to consider for the origin of AT~2015cf are gap transients such as intermediate luminosity red transients (ILRTs) and luminous red novae (LRNe), which are red in colour and intrinsically fainter than most CCSNe (e.g. \citealt{pastorello2019}).

The ILRT events have been suggested in the literature to be SNe where electron capture reactions of $^{24}$Mg or $^{20}$Ne in the O-Ne-Mg core of a low-mass massive star would trigger a terminal SN explosion rather than the core-collapse of an iron core (e.g. \citealt{cai2021}). SN 2008S (e.g. \citealt{prieto2008,botticella2009,wanajo2009}) is a prototypical member of these electron-capture SN candidates. The LRN transients have been associated in the literature with stellar merger powered common envelope events in close binary systems (e.g. \citealt{pastorello2019b}); thus, these events are likely not terminal explosions. These events are characterised by double-peaked light curves and are thought to be more massive versions of the classical Galactic red nova V1309 Sco \citep{tylenda2011}. While fainter than typical CCSNe, the volumetric rate of SN~2008S-like ILRTs could be relatively high compared to the CCSN rate (up to $\sim$10~\%; \citealt{mattila2012}). The rate of LRNe is more uncertain.

\begin{figure*}
\centering
\includegraphics[width=0.33\linewidth]{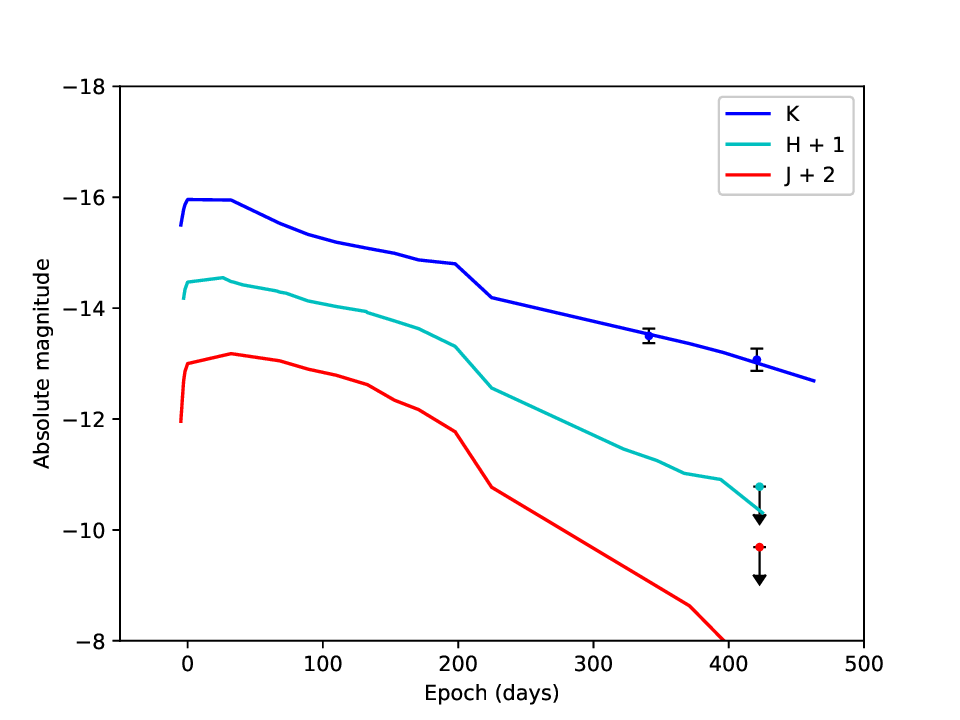}
\includegraphics[width=0.33\linewidth]{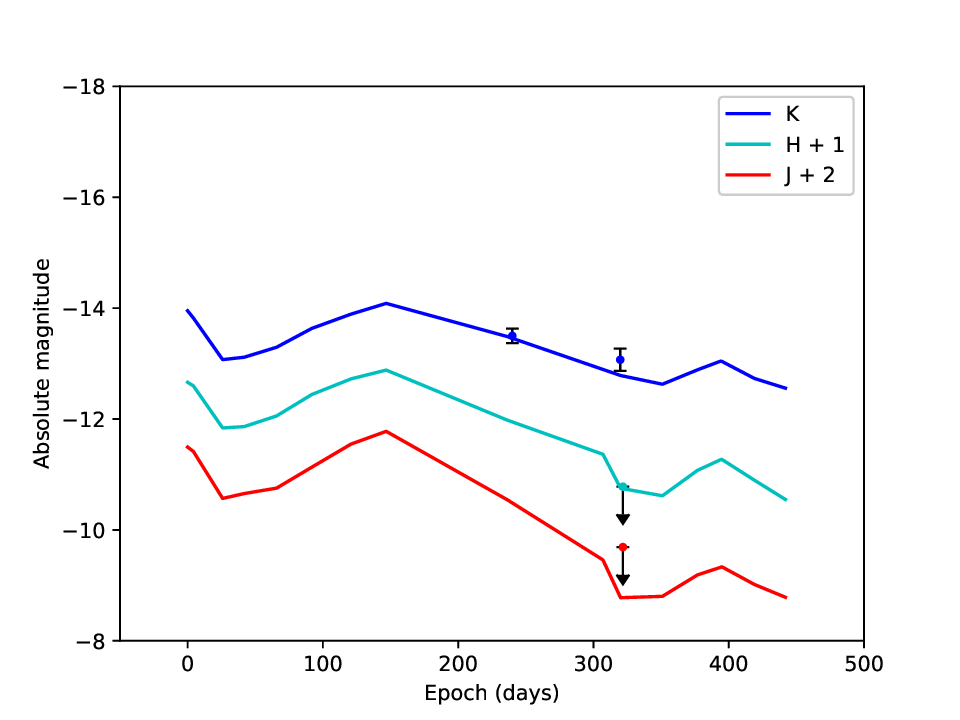}
\includegraphics[width=0.33\linewidth]{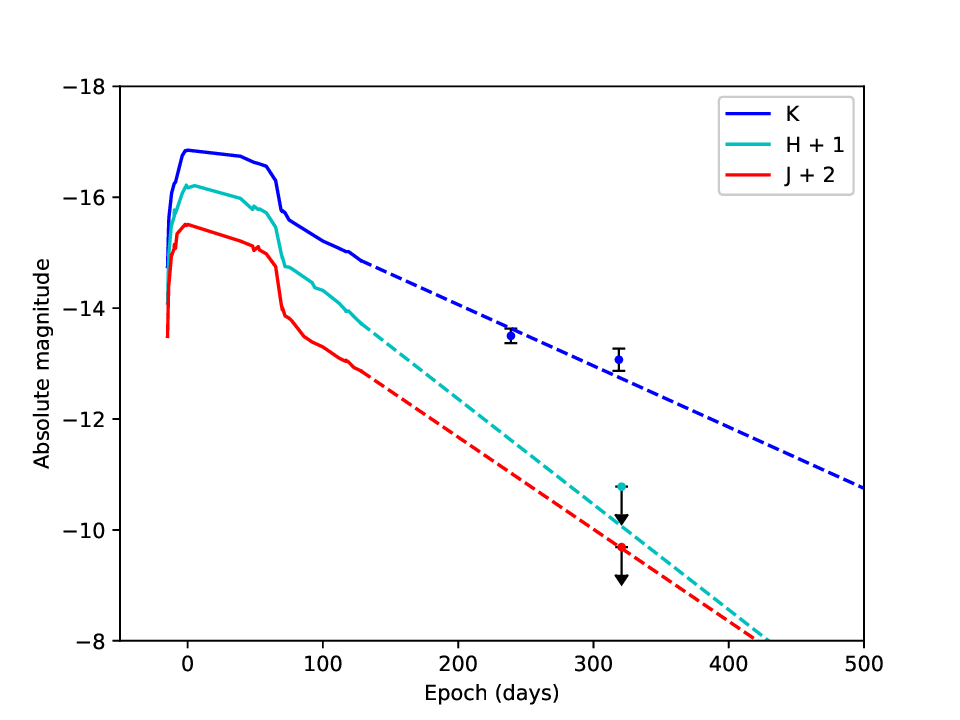}
\caption{Absolute light curves of AT~2015cf (with detections as points and limits with downwards pointing arrows) compared to a selection of events (curves) including the ILRT AT 2019abn (left), the LRN AT 2021blu (centre), and the Type IIP SN 2023ixf (right). The extrapolated tail phase of SN~2023ixf is indicated with dotted lines. The intrinsic brightness of the events has not been adjusted and no host galaxy extinction is assumed for AT~2015cf.}
\label{at_comparisons}
\end{figure*}

Unfortunately, well sampled late-time \textit{JHK}-band light curves of CCSNe are rare in the literature. \cite{kool2018} carried out a comparison of the observations of AT 2015cf to the near-IR evolution of representative examples of H-rich and H-poor SNe of the Type IIP SN 1999em \citep{krisciunas2009} and the Type IIb SN 2011dh \citep{ergon2014,ergon2015}, respectively. With adjustments to the intrinsic brightness of the comparison events, the observational epoch, and the host galaxy extinction, contextual similarity to the observations of AT 2015cf can be achieved; however, in both cases the \textit{K}-band brightness of these template events declines more rapidly than the observations of AT 2015cf. While the nature of AT 2015cf remained uncertain, \cite{kool2018} favoured the H-rich CCSN origin for the event. Recently, late-time \textit{JHK}-band observations have been reported in the literature for a variety of transients, which enabled us to carry out new $\chi^2$ comparisons to ILRTs, LRNe, and CCSNe. We highlight comparisons to the ILRT AT~2019abn ($\mu = 29.67$~mag, $A_V^{\mathrm{tot}} = 1.936$~mag; \citealt{valerin2025}), the LRN AT~2021blu ($\mu = 29.68$~mag, $A_V^{\mathrm{tot}} = 0.062$~mag; \citealt{pastorello2023}), and the Type IIP SN~2023ixf ($\mu = 29.17$~mag, $A_V^{\mathrm{tot}} = 0.121$~mag; \citealt{singh2024}) in Fig.~\ref{at_comparisons}. In the case of SN 2023ixf the \textit{JHK}-band light curves do not extend further than roughly 128 d from the approximate \textit{K}-band peak; therefore, we linearly extrapolated the tail phase evolution with a linear fit. In all the comparisons, we do not adjust the intrinsic brightness of the events, and we assume no host extinction for AT 2015cf. In all these cases reasonable matches can be achieved assuming that the first detection of AT 2015cf took place within a 200 to 400 d range from the onset of the event. If AT 2015cf was reddened by some host galaxy extinction similar light curve matches would have been possible if AT 2015cf was detected somewhat earlier and/or the comparison events would have been intrinsically somewhat brighter. However, we note that AT 2019abn and AT 2021blu are located already in the bright end of the observed ILRT and LRN distributions, respectively.

Based on these comparisons, we favour the simplest explanation that AT 2015cf is a terminal SN explosion of an unconfirmed type. However, for completeness, we report the results for the missing and undetectable fractions for both cases of including and excluding AT 2015cf from the sample of SNe discovered in the GSAOI/GeMS monitoring programme of LIRGs; this does not have a major effect on our conclusions when the other uncertainties are taken into account.

\section{Results}
\label{results}

\begin{figure*}[!h]
\centering
\includegraphics[width=.45\linewidth]{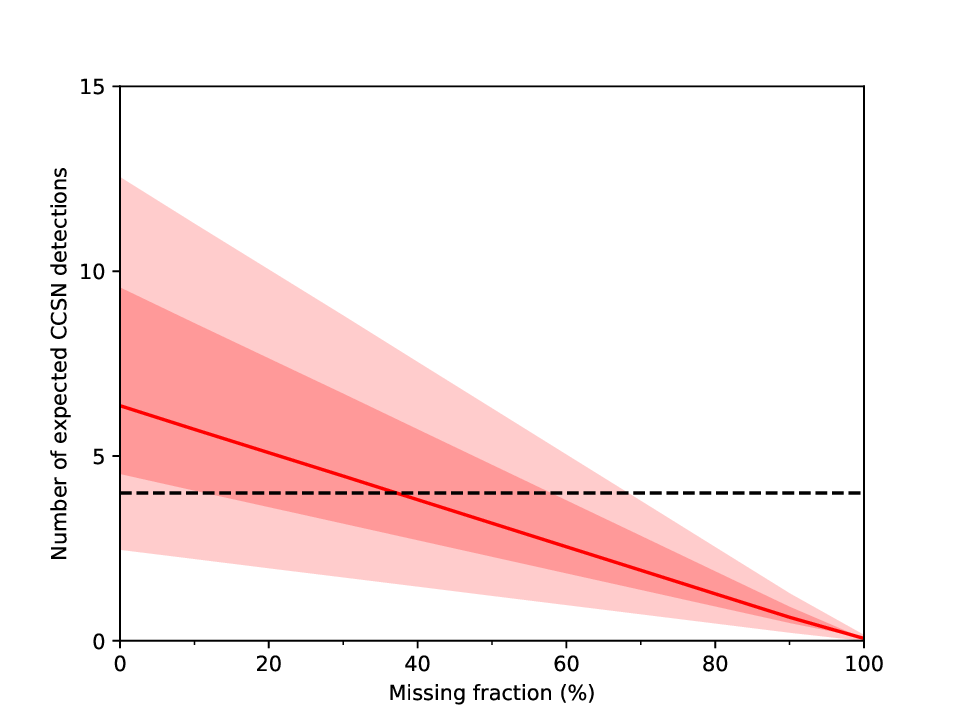}
\includegraphics[width=.45\linewidth]{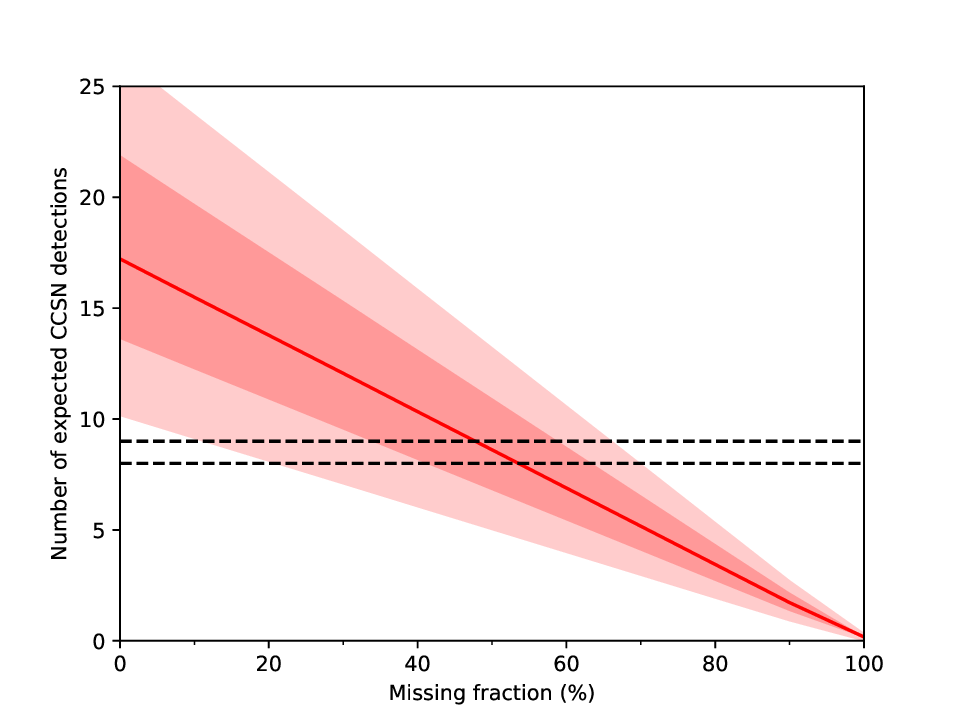}
\caption{Number of expected CCSN detections over the survey period of the dataset as a function of the missing fraction. Left: The GSAOI/GeMS dataset. Right: The GSAOI/GeMS dataset combined with the ALTAIR/NIRI dataset from Paper I. Red line: The mean value. Shaded red areas: 1$\sigma$ and 2$\sigma$ confidence intervals. Poissonian upper and lower limits due to small number statistics and the cumulative effect of other error sources are combined here. Dashed black: Number of real detected CCSNe in the survey(s). We note that while the expected CCSN counts for the datasets are 9 and 25 CCSNe for GSAOI/GeMS alone, and GSAOI/GeMS combined with ALTAIR/NIRI dataset, respectively, the Monte Carlo simulations result in detecting significantly less CCSNe even at 0 \% missing fraction. This is caused by a combined effect of faint CCSNe, moderate extinction and long time interval until the next observation epoch within the yielded CT.}
\label{missing_fracs}
\end{figure*}

    The output of a single run of the Monte Carlo simulation is an average detection probability for any CCSN subtype, for a given missing fraction and LIRG. We calculated the expected number of CCSNe in the dataset over the survey period (which excludes CCSNe that occur during survey gaps) based on the TCTs of the dataset and the intrinsic CCSN rates inferred based on the SED MCMC template fitting of the sample galaxies. By running the simulation with different missing fraction values we explored how the resulting estimated CCSN detection probabilities compare with the number of detected CCSNe. By multiplying the estimated intrinsic number of CCSNe with the different CCSN detection probabilities from the Monte Carlo simulation, we calculated the expected number of CCSN detections for each explored missing fraction value. Finally, we compared the number of expected CCSN detections to the number of real detections to constrain the missing fraction required to explain the discrepancy (see Fig. \ref{missing_fracs}). 

    Because the CCSN nature of event AT 2015cf is uncertain, we calculate the missing fraction for both cases. Assuming that AT 2015cf was a genuine CCSN, based on the number of four detected CCSNe in the dataset, we obtain a missing fraction of $37.2^{+21.2}_{-26.5}$\%. However, excluding the event AT 2015cf results in a missing fraction of $52.8^{+15.9}_{-19.8}$\%. 
    
    Because the method used in Paper I is the same as in this work, we can reduce the effect of small number statistics  by combining the results of this work with the results obtained in Paper I with Near-InfraRed Imager (NIRI; \citealt{hodapp2003}) and the ALTtitude conjugate Adaptive optics for the InfraRed (ALTAIR). The combined dataset increased the number of expected CCSNe in the dataset from 16 to 25, reducing the Poissonian small number statistical error by $\sim20\%$ The combined dataset contains nine or eight (excluding AT 2015cf) detected CCSNe, which results in a missing fraction of $47.8^{+11.4}_{-14.4}$\% or $53.6^{+10.2}_{-12.9}$\%, respectively. We note that because the width of the confidence intervals are dependent on the value of the missing fraction, the improvement in the statistics might not seem apparent.
    
    The missing fraction of CCSNe in LIRGs represents the fraction of events that are undetectable by all optical and near-IR surveys due to the extreme host extinctions. In addition to the missing fraction, CCSNe remain undetected due to more moderate host extinctions. We estimate that fraction by integrating the host extinction model (Sect \ref{host_extinction}) from infinity to a set of extinction thresholds. For example, assuming an optically discovered CCSN with the highest host extinction from \cite{kankare2021} as the extinction threshold with $A_V \approx 3$ mag, integrating the host extinction results in 78 \% remaining undetected within the surface extinction distribution. After this fraction is applied to the population of CCSNe not part of the missing fraction, we obtain a total undetectable fraction of $88.3^{+2.6}_{-3.2}$\% based on the combined dataset with a total of nine CCSNe. The highest host extinction in the CCSN sample used to determine the host extinction model is $A_V = 16$ mag. Taking this as an extinction threshold for \textit{K}-band surveys, we obtain a total undetectable fraction of $61.4^{+8.5}_{-10.6}$\%. 
    The total undetectable fractions for a sample of extinction thresholds and both datasets are presented in Table \ref{host_ext_missing_fracion}.

    Since the results between Paper I and this work are very similar, comparisons to previous work done in Paper I also apply here, summarised below. \cite{miluzio2013} found a total fraction of $\sim 60\%$ to $75\%$ of CCSNe in LIRGs to remain undiscovered in \textit{K}-band. Comparing this to our total undetectable fraction of \textit{K}-band surveys we find them in good agreement. This is unsurprising since the observed wavelength and sample of galaxies are similar.
    
    \cite{mattila2012} found a missing fraction of $83^{+9}_{-15}$\% for LIRGs with optical surveys assuming an extinction correction similar to that of normal galaxies; therefore, this result roughly corresponds to our derived undetectable fraction. Our estimate of undetectable fraction for optical surveys using $A_V = 3$ mag as the threshold is well consistent with this value within errors; however, we do note that due to the error ranges we find agreeable undetectable fractions for threshold values between $A_V = 2$ mag to $A_V \approx 10$ mag. Their result was based on a single frequently observed LIRG, Arp 299.

    For completeness, we report the median CT of our fastest evolving template (Type IIb) as 109 days. Therefore, for a similar survey with a cadence of $\sim$100 days, one expects to have at least one detection for 50\% of the detectable Type IIb SNe. 

\begin{table}[!h]
\begin{center}
\caption{Total undetectable fractions calculated from the missing fraction and the extinction model.}
\begin{tabular}{ccccccc}
\hline
\hline
$A_{V,\mathrm{limit}}$\tablefootmark{a} & $f_{A_V}$\tablefootmark{b} & $f_{\mathrm{tot-3}}$\tablefootmark{c} & $f_{\mathrm{tot-4}}$\tablefootmark{d} & $f_{\mathrm{tot-8}}$\tablefootmark{e} & $f_{\mathrm{tot-9}}$\tablefootmark{f}\\
(mag) & (\%) & (\%) & (\%) & (\%) & (\%) \\
\hline
2 & 15.5 & $92.7^{+2.5 }_{-3.1 }$ & $90.3^{+3.3 }_{-4.1 }$ & $92.8^{+1.6}_{-2.0 }$ & $91.9^{+1.8 }_{-2.2 }$\\
3 & 22.3 & $89.5^{+3.5 }_{-4.4 }$ & $86.0^{+4.7 }_{-5.9 }$ & $89.6^{+2.3}_{-2.9 }$ & $88.3^{+2.6 }_{-3.2 }$\\
4 & 28.5 & $86.5^{+4.5 }_{-5.6 }$ & $82.1^{+6.0 }_{-7.6 }$ & $86.8^{+2.9}_{-3.7 }$ & $85.1^{+3.3 }_{-4.1 }$\\
5 & 34.3 & $83.8^{+5.5 }_{-6.8 }$ & $78.4^{+7.3 }_{-9.1 }$ & $84.1^{+3.5}_{-4.4 }$ & $82.1^{+3.9 }_{-4.9 }$\\
7 & 44.5 & $79.0^{+7.1 }_{-8.8 }$ & $72.0^{+9.4 }_{-11.8}$ & $79.3^{+4.5}_{-5.7 }$ & $76.7^{+5.1 }_{-6.4 }$\\
10& 56.8 & $73.2^{+9.0 }_{-11.3}$ & $64.3^{+12.0}_{-15.1}$ & $73.6^{+5.8}_{-7.3 }$ & $70.3^{+6.5 }_{-8.2 }$\\
13& 66.4 & $68.7^{+10.6}_{-13.2}$ & $58.3^{+14.1}_{-17.6}$ & $69.2^{+6.8}_{-8.6 }$ & $65.3^{+7.6 }_{-9.6 }$\\
16& 73.9 & $65.1^{+11.8}_{-14.6}$ & $53.6^{+15.6}_{-19.6}$ & $65.7^{+7.5}_{-9.5 }$ & $61.4^{+8.5 }_{-10.6}$\\
20& 81.4 & $61.6^{+13.0}_{-16.1}$ & $48.8^{+17.2}_{-21.6}$ & $62.2^{+8.3}_{-10.5}$ & $57.5^{+9.3 }_{-11.7}$\\
25& 87.7 & $58.6^{+14.0}_{-17.4}$ & $44.9^{+18.6}_{-23.2}$ & $59.3^{+8.9}_{-11.3}$ & $54.2^{+10.0}_{-12.6}$\\
30& 92.0 & $56.6^{+14.6}_{-18.2}$ & $42.2^{+19.5}_{-24.4}$ & $57.3^{+9.4}_{-11.9}$ & $51.9^{+10.5}_{-13.2}$\\
40& 96.4 & $54.5^{+15.3}_{-19.1}$ & $39.4^{+20.4}_{-25.6}$ & $55.2^{+9.8}_{-12.4}$ & $49.6^{+11.0}_{-13.9}$\\
\hline
\end{tabular}
\label{host_ext_missing_fracion}
\tablefoot{\tablefoottext{a}{The extinction threshold, } \tablefoottext{b}{the fraction of CCSNe with host extinctions below the extinction threshold within the extinction correction, }$^{\mathrm{(c)-(f)}}$~the total undetectable fraction assuming a total of three, four, eight, and nine detected CCSNe, respectively.}
\end{center}
\end{table}

\section{Discussion}
\label{discussion}

    The results derived in this work are based on LIRGs in the local Universe (median distance 84 Mpc; $z$ = 0.018), and applying them to LIRGs at larger redshifts is not straightforward, in particular due to the evolution of this galaxy population. \cite{magnelli2011} measured the contributions of normal galaxies and (U)LIRGs to the cosmic SFR density at $0.0 < z < 2.3$, and found that LIRGs account for roughly 5, 35, 50, and 50 \% of the SFR density at redshifts of 0.0, 0.5, 1.0, and 2.0, respectively. The SFR density is traced by the CCSN rate, and assuming that our result of $\sim90$ \% undetectable fraction of CCSNe applies for LIRGs at all redshifts when probing the rest frame optical wavelengths, a major fraction of CCSNe remain undiscovered regardless of the survey depth beyond the local Universe. Thus, the undetectable fraction of CCSNe in LIRGs has a crucial effect on the observed cosmic CCSN rates and their usage as independent probes of the cosmic star formation history beyond the local Universe (see e.g. \citealt{decoursey2025}). Furthermore, \cite{mattila2012} estimate that nearly 100 \% of CCSNe in the local ULIRGs are not detected by optical surveys. The contribution of ULIRGs to the SFR density increases with redshift, reaching $\sim20$ \% at $z$ = 2 (\citealt{magnelli2011}).

\section{Conclusions}
\label{conclusions}

    We have presented updated estimates for the fraction of CCSNe in LIRGs in the local Universe that remain undetected by SN surveys, with a detailed description of our method. The analysis in this work is based on the SUNBIRD survey near-IR $K$-band monitoring dataset of nine LIRGs using the Gemini-South telescope with the GSAOI/GeMS AO system. The survey detected four transients, of which three are photometrically classified as CCSNe, and one is tentatively a CCSN discovered at late times.

    We present the total undetectable fractions based on the analysed dataset, which consist of the missing fraction, and an additional survey extinction threshold dependent fraction of CCSNe that remain undiscovered due to moderate extinctions. We present these for a selection of survey extinction thresholds, and highlight two cases of $A_V$ = 3 and 16 mag, which act as proxies for optical and near-IR surveys, respectively. Based on four CCSNe in the dataset, we find total undetectable fractions of $86.0^{+4.7}_{-5.9}$\% and $53.6^{+15.6}_{-19.6}$\% for optical and near-IR surveys, respectively. 

    Finally, we present the aggregate results of this work combined with those from Paper I derived with the same method for the SUNBIRD ALTAIR/NIRI laser guide star AO dataset of a sample of nine LIRGs monitored using the Gemini-North telescope. We find a total undetectable fractions of $88.3^{+2.6}_{-3.2}$\% and $61.4^{+8.5}_{-10.6}$\% for optical and near-IR surveys, respectively. The facilities such as the Vera C. Rubin Observatory and the James Webb Space Telescope will probe the intrinsic CCSN rates with an unprecedented accuracy, on which the undetectable fraction of CCSNe will have a crucial impact beyond the local Universe at higher redshifts dominated by (U)LIRGs.

\begin{acknowledgements}

    We thank the anonymous referee for useful comments.

    IM, EK, and KM acknowledge financial support from the Emil Aaltonen foundation.

    TMR is part of the Cosmic Dawn Center (DAWN), which is funded by the Danish National Research Foundation under grant DNRF140. TMR and SM acknowledge support from the Research Council of Finland project 350458.
    
    Based on observations obtained at the international Gemini Observatory, a program of NSF NOIRLab, which is managed by the Association of Universities for Research in Astronomy (AURA) under a cooperative agreement with the U.S. National Science Foundation on behalf of the Gemini Observatory partnership: the U.S. National Science Foundation (United States), National Research Council (Canada), Agencia Nacional de Investigaci\'{o}n y Desarrollo (Chile), Ministerio de Ciencia, Tecnolog\'{i}a e Innovaci\'{o}n (Argentina), Minist\'{e}rio da Ci\^{e}ncia, Tecnologia, Inova\c{c}\~{o}es e Comunica\c{c}\~{o}es (Brazil), and Korea Astronomy and Space Science Institute (Republic of Korea).

    Based on observations made with the Nordic Optical Telescope, owned in collaboration by the University of Turku and Aarhus University, and operated jointly by Aarhus University, the University of Turku and the University of Oslo, representing Denmark, Finland and Norway, the University of Iceland and Stockholm University at the Observatorio del Roque de los Muchachos, La Palma, Spain, of the Instituto de Astrofisica de Canarias.
    
    Observations from the NOT were obtained through the NUTS2 collaboration which is supported in part by the Instrument Centre for Danish Astrophysics (IDA), and the Finnish Centre for Astronomy with ESO (FINCA) via Academy of Finland grant nr 306531.

    This publication makes use of data products from the Two Micron All Sky Survey, which is a joint project of the University of Massachusetts and the Infrared Processing and Analysis Center/California Institute of Technology, funded by the National Aeronautics and Space Administration and the National Science Foundation.

    This research has made use of the NASA/IPAC Extragalactic Database (NED), which is operated by the Jet Propulsion Laboratory, California Institute of Technology, under contract with the National Aeronautics and Space Administration.

    This research has made use of the NASA/IPAC Infrared Science Archive, which is funded by the National Aeronautics and Space Administration and operated by the California Institute of Technology.
\end{acknowledgements}

\bibliography{sources}

\begin{appendix}

\section{Additional tables}

\begin{table}[h]
\centering
\caption{Survey epochs and CTs of ESO 264-G036.}
    \resizebox{0.49\textwidth}{!}{%
\begin{tabular}{cccccccccc}
\hline
\hline
MJD & $m_{\mathrm{lim}}$ & $\Delta t$ & IIb & Ib & Ic & II & 87A & IIn & Program ID\\
 & (mag) &  & (d) & (d) & (d) & (d) & (d) & (d) & \\
\hline
56671 & 20.80 &  & 105 & 112 & 120 & 146 & 185 & 667 & 2013B-Q-65\\
57005 & 20.89 & 334 & 108 & 116 & 125 & 151 & 191 & 334 & 2014B-C-1\\
57088 & 20.80 & 83 & 83 & 83 & 83 & 83 & 83 & 83 &  2015A-C-2\\
57437 & 20.30 & 349 & 84 & 92 & 100 & 117 & 150 & 349 &  2016A-C-1\\
\hline
\end{tabular}}
\label{ct_ESO_264-G036}
\end{table}

\begin{table}[h]
\caption{Survey epochs and CTs of ESO 267-G030.}
    \resizebox{0.49\textwidth}{!}{%
\begin{tabular}{cccccccccc}
\hline
\hline
MJD & $m_{\mathrm{lim}}$ & $\Delta t$ & IIb & Ib & Ic & II & 87A & IIn & Program ID\\
 & (mag) &  & (d) & (d) & (d) & (d) & (d) & (d) & \\
 \hline
56435 & 21.06 &  & 118 & 126 & 134 & 169 & 209 & 766 & 2013A-Q-9\\
57088 & 21.06 & 653 & 119 & 127 & 135 & 168 & 209 & 653 & 2015A-C-2\\
57174 & 20.87 & 86 & 86 & 86 & 86 & 86 & 86 & 86 & 2015A-Q-7\\
57437 & 20.08 & 263 & 79 & 85 & 95 & 110 & 144 & 263 & 2016A-C-1\\
\hline
\end{tabular}}
\end{table}

\begin{table}[h]
\caption{Survey epochs and CTs of ESO 440-IG058.}
    \resizebox{0.49\textwidth}{!}{%
\begin{tabular}{cccccccccc}
\hline
\hline
MJD & $m_{\mathrm{lim}}$ & $\Delta t$ & IIb & Ib & Ic & II & 87A & IIn & Program ID\\
 & (mag) &  & (d) & (d) & (d) & (d) & (d) & (d) & \\
 \hline
56321 & 20.32 &  & 77 & 84 & 91 & 106 & 137 & 472 & 2012B-SV-407\\
56401 & 20.35 & 80 & 78 & 80 & 80 & 80 & 80 & 80 & 2013A-Q-9\\
57086 & 20.35 & 685 & 78 & 86 & 93 & 107 & 141 & 482 & 2015A-Q-7\\
57437 & 19.42 & 351 & 44 & 50 & 58 & 65 & 83 & 248 & 2016A-C-1\\
\hline
\end{tabular}}
\end{table}

\begin{table}[h]
\caption{Survey epochs and CTs of IRAS 08355-4944.}
    \resizebox{0.49\textwidth}{!}{%
\begin{tabular}{cccccccccc}
\hline
\hline
MJD & $m_{\mathrm{lim}}$ & $\Delta t$ & IIb & Ib & Ic & II & 87A & IIn & Program ID\\
 & (mag) &  & (d) & (d) & (d) & (d) & (d) & (d) & \\
 \hline
57086 & 19.66 &  & 42 & 47 & 54 & 62 & 78 & 235 & 2015A-Q-7\\
57442 & 19.66 & 356 & 42 & 47 & 54 & 62 & 78 & 235 & 2016A-C-1\\
\hline
\end{tabular}}
\end{table}

\begin{table}[h]
\caption{Survey epochs and CTs of IRAS 17138-1017.}
    \resizebox{0.49\textwidth}{!}{%
\begin{tabular}{cccccccccc}
\hline
\hline
MJD & $m_{\mathrm{lim}}$ & $\Delta t$ & IIb & Ib & Ic & II & 87A & IIn & Program ID\\
 & (mag) &  & (d) & (d) & (d) & (d) & (d) & (d) & \\
 \hline
56373 & 21.46 &  & 141 & 151 & 158 & 201 & 247 & 925 & 2013A-Q-9\\
56435 & 20.37 & 62 & 62 & 62 & 62 & 62 & 62 & 62 & 2013A-Q-9\\
56759 & 21.20 & 324 & 131 & 139 & 147 & 182 & 231 & 324 & 2014A-Q-21\\
56762 & 21.46 & 3 & 3 & 3 & 3 & 3 & 3 & 3 & 2014A-Q-21\\
57087 & 20.37 & 325 & 96 & 105 & 112 & 134 & 171 & 325 & 2015A-C-2\\
57174 & 21.20 & 87 & 87 & 87 & 87 & 87 & 87 & 87 & 2015A-Q-7\\
\hline
\end{tabular}}
\end{table}

\begin{table}[h]
\caption{Survey epochs and CTs of IRAS 18293-3413.}
    \resizebox{0.49\textwidth}{!}{%
\begin{tabular}{cccccccccc}
\hline
\hline
MJD & $m_{\mathrm{lim}}$ & $\Delta t$ & IIb & Ib & Ic & II & 87A & IIn & Program ID\\
 & (mag) &  & (d) & (d) & (d) & (d) & (d) & (d) & \\
 \hline
56403 & 20.61 &  & 111 & 120 & 127 & 157 & 197 & 721 & 2013A-Q-9\\
56436 & 20.62 & 33 & 33 & 33 & 33 & 33 & 33 & 33 & 2013A-Q-9\\
56454 & 19.97 & 18 & 18 & 18 & 18 & 18 & 18 & 18 & 2013A-Q-9\\
57175 & 20.62 & 721 & 111 & 119 & 126 & 157 & 197 & 720 & 2015A-Q-6\\
\hline
\end{tabular}}
\end{table}

\begin{table}[h]
\caption{Survey epochs and CTs of NGC 3110.}
    \resizebox{0.49\textwidth}{!}{%
\begin{tabular}{cccccccccc}
\hline
\hline
MJD & $m_{\mathrm{lim}}$ & $\Delta t$ & IIb & Ib & Ic & II & 87A & IIn & Program ID\\
 & (mag) &  & (d) & (d) & (d) & (d) & (d) & (d) & \\
 \hline
57092 & 20.56 &  & 116 & 126 & 133 & 165 & 207 & 754 & 2015A-Q-7\\
57172 & 19.82 & 80 & 80 & 80 & 80 & 80 & 80 & 80 & 2015A-Q-7\\
57437 & 20.56 & 265 & 115 & 125 & 133 & 166 & 208 & 265 & 2016A-C-1\\
57500 & 19.82 & 63 & 63 & 63 & 63 & 63 & 63 & 63 & 2016A-C-2\\
\hline
\end{tabular}}
\end{table}

\begin{table}[h]
\caption{Survey epochs and CTs of NGC 3508.}
    \resizebox{0.49\textwidth}{!}{%
\begin{tabular}{cccccccccc}
\hline
\hline
MJD & $m_{\mathrm{lim}}$ & $\Delta t$ & IIb & Ib & Ic & II & 87A & IIn & Program ID\\
 & (mag) &  & (d) & (d) & (d) & (d) & (d) & (d) & \\
 \hline
57088 & 20.72 &  & 145 & 155 & 162 & 206 & 255 & 961 & 2015A-C-2\\
57440 & 20.72 & 352 & 145 & 155 & 162 & 206 & 255 & 352 & 2016A-C-1\\
\hline
\end{tabular}}
\end{table}

\begin{table}[h]
\caption{Survey epochs and CTs of NGC 4575.}
    \resizebox{0.49\textwidth}{!}{%
\begin{tabular}{cccccccccc}
\hline
\hline
MJD & $m_{\mathrm{lim}}$ & $\Delta t$ & IIb & Ib & Ic & II & 87A & IIn & Program ID\\
 & (mag) &  & (d) & (d) & (d) & (d) & (d) & (d) & \\
 \hline
57087 & 20.05 &  & 115 & 124 & 131 & 161 & 203 & 741 & 2015A-C-2\\
57440 & 20.05 & 353 & 115 & 123 & 132 & 162 & 202 & 353 & 2016A-C-1\\
\hline
\end{tabular}}
\label{ct_ngc_4575}
\end{table}

\end{appendix}

\end{document}